\documentclass[twocolumn]{aastex701}
\usepackage{placeins}

\usepackage{amsmath}
\usepackage{graphicx}
\usepackage{booktabs}
\usepackage{array}
\usepackage{hyperref}

\graphicspath{{figures/}}

\newcommand{\Teff}{T_{\rm eff}}
\newcommand{\logg}{\log g}
\newcommand{\mh}{[\mathrm{M/H}]}

\newcommand{\rjup}{R_{\rm Jup}}
\newcommand{\mjup}{M_{\rm Jup}}

\shorttitle{COSMOS-3D: Shabgard}
\shortauthors{Salmasi et al.}

\begin{document}

\title{COSMOS-3D: Spectro-photometric Confirmation of Shabgard, a Cool, Metal-poor Brown Dwarf in the COSMOS Field}

\correspondingauthor{Bahareh S. Salmasi}

\author[orcid=0009-0003-0031-0527]{Bahareh S. Salmasi}
\affiliation{Institute of Astronomy, National Tsing Hua University, Hsinchu, Taiwan}
\affiliation{School of Mathematical and Physical Sciences, Macquarie University, Sydney, NSW 2109, Australia}
\affiliation{Astrophysics and Space Technologies Research Centre, Macquarie University, Sydney, NSW 2109, Australia}
\email[show]{physics.bahare@gmail.com}

\author[orcid=0000-0002-0524-5328]{Priyanka Jalan}
\affiliation{Institute of Astronomy, National Tsing Hua University, Hsinchu, Taiwan}
\affiliation{Center for Theoretical Physics, Polish Academy of Sciences, Al. Lotników 32/46, 02-668, Warsaw, Poland}
\email{priyajalan14@gmail.com}

\author[orcid=0000-0002-6821-8669]{Tomotsugu Goto}
\affiliation{Institute of Astronomy, National Tsing Hua University, Hsinchu, Taiwan}
\affiliation{Department of Physics, National Tsing Hua University, Hsinchu, Taiwan}
\email{itlhnis@gmail.com}

\author[orcid=0000-0001-7228-1428]{Tetsuya Hashimoto}
\affiliation{Institute of Astronomy, National Chung Hsing University, Taichung, Taiwan}
\email{tetsuya@phys.nchu.edu.tw}

\author[orcid=0000-0002-7633-431X]{Feige Wang}
\affiliation{Department of Astronomy, University of Michigan, Ann Arbor, MI, USA}
\email{fgwang@umich.edu}

\author[orcid=0000-0003-3935-7018]{Tayyaba Zafar}
\affiliation{School of Mathematical and Physical Sciences, Macquarie University, Sydney, NSW 2109, Australia}
\affiliation{Astrophysics and Space Technologies Research Centre, Macquarie University, Sydney, NSW 2109, Australia}
\email{tayyaba.zafar@mq.edu.au}

\author[orcid=0000-0001-6874-1321]{Koki Kakiichi}
\affiliation{Cosmic Dawn Center (DAWN), Niels Bohr Institute,
University of Copenhagen, Jagtvej 128, DK-2200 Copenhagen N, Denmark}
\email{koki.kakiichi@nbi.ku.dk}

\author[orcid=0000-0002-9119-2313]{Ece Kilerci}
\affiliation{Department of Astronomy and Space Sciences, Science Faculty, \.{I}stanbul University, Beyaz{\i}t 34119, \.Istanbul, T\"{u}rkiye}
\email{ecekilerci@gmail.com }

\author[orcid=0009-0002-7530-0019]{Srivardini Ayyappan}
\affiliation{Institute of Astronomy, National Tsing Hua University, Hsinchu, Taiwan}
\affiliation{School of Mathematical and Physical Sciences, Macquarie University, Sydney, NSW 2109, Australia}
\affiliation{Astrophysics and Space Technologies Research Centre, Macquarie University, Sydney, NSW 2109, Australia}
\email{srivardini.ayyappan@hdr.mq.edu.au}

\author[orcid=0000-0001-5492-4522]{Romain A. Meyer}
\affiliation{Department of Astronomy, University of Geneva,
Chemin Pegasi 51, 1290 Versoix, Switzerland}
\email{romain.meyer@unige.ch}

\author[orcid=0009-0003-1221-1630]{Anilkumar Mailvaganam}
\affiliation{School of Mathematical and Physical Sciences, Macquarie University, Sydney, NSW 2109, Australia}
\affiliation{Astrophysics and Space Technologies Research Centre, Macquarie University, Sydney, NSW 2109, Australia}
\email{anilkumar.mailvaganam@hdr.mq.edu.au}

\author[orcid=0009-0004-7639-869X]{Tamal Mukherjee}
\affiliation{School of Mathematical and Physical Sciences, Macquarie University, Sydney, NSW 2109, Australia}
\affiliation{Astrophysics and Space Technologies Research Centre, Macquarie University, Sydney, NSW 2109, Australia}
\email{tamal.mukherjee@hdr.mq.edu.au}

\author[orcid=0009-0005-6537-9597]{Xuanyi Lyu}
\affiliation{School of Mathematical and Physical Sciences, Macquarie University, Sydney, NSW 2109, Australia}
\email{xuanyi.lyu@hdr.mq.edu.au}

\author[orcid=0009-0005-0489-6232]{Imam Uzma Aalia}
\affiliation{School of Mathematical and Physical Sciences, Macquarie University, Sydney, NSW 2109, Australia}
\affiliation{Astrophysics and Space Technologies Research Centre, Macquarie University, Sydney, NSW 2109, Australia}
\email{aaliaimam.uzma@hdr.mq.edu.au}

\author[orcid=0000-0001-5980-0174]{Arihant Raidani}
\affiliation{School of Mathematical and Physical Sciences, Macquarie University, Sydney, NSW 2109, Australia}
\affiliation{Astrophysics and Space Technologies Research Centre, Macquarie University, Sydney, NSW 2109, Australia}
\email{arihant.raidani@students.mq.edu.au}

\author[orcid=0009-0005-2766-2850]{Samuel Carlson}
\affiliation{School of Mathematical and Physical Sciences, Macquarie University, Sydney, NSW 2109, Australia}
\affiliation{Astrophysics and Space Technologies Research Centre, Macquarie University, Sydney, NSW 2109, Australia}
\email{samuel.carlson@students.mq.edu.au}

\author[orcid=0009-0000-9947-9805]{Jose Luis Carrillo Martinez}
\affiliation{School of Mathematical and Physical Sciences, Macquarie University, Sydney, NSW 2109, Australia}
\affiliation{Astrophysics and Space Technologies Research Centre, Macquarie University, Sydney, NSW 2109, Australia}
\email{joseluis.carrillomartinez@hdr.mq.edu.au}

\author[orcid=0000-0002-9382-9832]{Andreas L. Faisst}
\affiliation{IPAC, California Institute of Technology, 1200 E. California Blvd. Pasadena, CA 91125, USA}
\email{afaisst@caltech.edu}

\author[orcid=0000-0002-7087-0701]{Marko Shuntov}
\affiliation{Cosmic Dawn Center (DAWN), Niels Bohr Institute,
University of Copenhagen, Jagtvej 128, DK-2200 Copenhagen N, Denmark}
\affiliation{University of Geneva, 24 rue du G\'en\'eral-Dufour,
1211 Gen\`eve 4, Switzerland}
\email{marko.shuntov@nbi.ku.dk}

\author[orcid=0000-0003-3596-8794]{Hollis B. Akins}
\affiliation{Department of Astronomy, The University of Texas at Austin,
2515 Speedway Blvd Stop C1400, Austin, TX 78712, USA}
\email{hollis.akins@utexas.edu}

\author[orcid=0000-0001-5951-459X]{Zihao Li}
\affiliation{Cosmic Dawn Center (DAWN), Niels Bohr Institute,
University of Copenhagen, Jagtvej 128, DK-2200 Copenhagen N, Denmark}
\email{zihao.li@nbi.ku.dk}
\collaboration{all}{COSMOS-3D Collaboration}

\begin{abstract}
Brown dwarfs are unique laboratories for studying the physics of cool
atmospheres, the low-mass end of star formation, and the transition
between stars and giant planets. Their spectra encode the effects of
temperature, gravity, metallicity, clouds, and disequilibrium chemistry,
but identifying and characterizing the coldest brown dwarfs remains
challenging, particularly at faint magnitudes in deep extragalactic
surveys. We aim to establish the nature and atmospheric properties of
COSMOS2025 ID~639188, hereafter \emph{Shabgard}, a compact source
previously selected as a brown-dwarf candidate from COSMOS-Web
photometry on the basis of its point-source morphology and
T/Y-dwarf-like NIRCam colors. In this work, we characterize this source
using serendipitous JWST/NIRCam F444W wide-field slitless spectroscopy
from COSMOS-3D. We extract its local-background-subtracted,
profile-weighted $4.05$-$4.85~\mu{\rm m}$ spectrum, combine it with
COSMOS2025 NIRCam photometry, compare the data with six atmosphere-model
families, and measure its motion using multi-epoch NIRCam imaging.

The LOWZ atmosphere-grid fit, based on principal component analysis
(PCA) interpolation and Markov Chain Monte Carlo (MCMC) sampling, gives
$T_{\rm eff}=718.6^{+10.6}_{-10.2}$~K and
$[{\rm M/H}]=-1.00^{+0.07}_{-0.08}$, favoring a cool,
subsolar-metallicity atmosphere characteristic of the late-T/Y
transition.

A fixed-atmosphere molecular-opacity test strongly favors a contribution
from CH$_4$ to the observed $4$-$5~\mu{\rm m}$ spectral shape. The restricted wavelength coverage, however, prevents a molecule-by-molecule abundance determination.

Model-dependent physical estimates span approximately
$23$-$52~M_{\rm Jup}$ and $140$-$180$~pc. We measure a proper motion
of $|\mu|=89.2\pm0.4$~mas~yr$^{-1}$, while the LOWZ MCMC flux scaling
gives a distance of $176.2^{+18.4}_{-17.8}$~pc, supporting the
interpretation of Shabgard as a Galactic brown dwarf. Together, these
results support its classification as a cool, metal-poor brown dwarf
and demonstrate the ability of JWST deep-field surveys to discover and
characterize faint Galactic substellar objects.
\end{abstract}


\section{Introduction}\label{sec:intro}

Brown dwarfs are intermediate mass objects, between massive planets and the smallest hydrogen-burning stars. Their masses are generally described in terms of the deuterium burning limit at $\sim13\mjup$ and the hydrogen burning limit at $\sim75$-$80\mjup$ \citep{Burrows2001}. However, both limitations rely on composition, helium percentage, and evolutionary history \citep{Chabrier2000, Spiegel2011}. Because brown dwarfs do not sustain stable core hydrogen fusion, they cool and fade continuously with time. Their spectra therefore trace a sequence of decreasing effective temperature through late-M, L, T, and Y types, with strong sensitivity to condensate clouds, collision-induced H$_2$ absorption, H$_2$O, CH$_4$, CO, CO$_2$, NH$_3$, PH$_3$, and other molecular species \citep{Kirkpatrick2005,Burrows2001,Marley2021,Phillips2020,Mukherjee2024}.

Brown dwarfs are important for both stellar and planetary astrophysics. They probe the low-mass end of star formation and the Galactic mass function \citep{Luhman2012,Kirkpatrick2021}, and their atmospheres provide directly observable analogs of self-luminous giant exoplanets \citep{MarleyRobinson2015,Faherty2016,Bowler2016}. Their spectra respond to $\Teff$, gravity, metallicity, clouds, and disequilibrium chemistry, so well-characterized brown dwarfs are essential benchmarks for atmospheric models used across the sub-stellar regime \citep{Marley2021,Karalidi2021,Mukherjee2024}. Old, metal-poor ultracool dwarfs are especially valuable because they can be used to study atmospheric physics in thick-disk and halo chemical environments where changes in opacity and collision-induced absorption alter colors and spectral shape \citep{Juric2008,Zhang2019,Meisner2021LOWZ,Burgasser2024}.

In deep extragalactic surveys, they are also a practical source-classification issue: cool Galactic dwarfs can be compact and red enough to contaminate high-redshift galaxy and active-galactic-nucleus samples \citep{Hainline2024,Burgasser2024,Chen2025,Tu2025}.
The James Webb Space Telescope (JWST; \citealt{Gardner2023JWST}) has substantially expanded the accessible volume for ultracool dwarf searches \citep{Wang2023TDwarf, Hainline2024}. Its near-infrared sensitivity, together with low-resolution spectroscopy from NIRSpec/PRISM, covers the wavelength range where late-T and Y dwarfs show strong molecular absorption and continuum-shape diagnostics \citep{Rigby2023JWST,Rieke2023NIRCam,Jakobsen2022NIRSpec}. NIRSpec/PRISM observations have already confirmed cold brown dwarfs at kiloparsec-scale distances in deep extragalactic fields, including objects identified in the Ultradeep NIRSpec and NIRCam Observations before the Epoch of Reionization (UNCOVER) \citep{Bezanson2024UNCOVER} Treasury program and the Red Unknowns: Bright Infrared Extragalactic Survey (RUBIES) \citep{Burgasser2024}. These detections show that JWST can extend spectroscopic studies of ultracool dwarfs to much larger distances than were previously accessible. However, interpreting such faint spectra remains challenging because the near-infrared continuum shape and broad molecular absorption features depend simultaneously on $\Teff$, metallicity, surface gravity, cloud opacity, and vertical mixing \citep{Burgasser2024,Mukherjee2024,Morrissey2026RUBIESBD,Tu2025}. Breaking these degeneracies typically requires both spectroscopy and external flux constraints, especially when only a limited wavelength range is available.

The COSMOS field provides a useful setting for our search because it combines deep multi-wavelength imaging with new wide-area JWST observations. In particular, COSMOS-Web supplies the photometric data for identifying compact red brown-dwarf candidates, while COSMOS-3D adds NIRCam/F444W wide-field slitless spectroscopy \citep{Casey2023,Shuntov2025,Meyer2026COSMOS3D}. Although the F444W grism window is narrower than a full NIRSpec/PRISM spectrum, it covers the $4$-$5\mu\mathrm{m}$ region where cool T/Y-like atmospheres are shaped by molecular opacity from species such as H$_2$O, CH$_4$, CO, and CO$_2$ \citep{Burgasser2024,Mukherjee2024}. It therefore provides a useful spectroscopic test of photometrically selected ultracool-dwarf candidates when combined with broadband JWST photometry.

In this work, we use COSMOS-Web photometry and COSMOS-3D NIRCam/F444W WFSS spectroscopy to test a photometrically selected brown-dwarf candidate in the COSMOS field. We start from the COSMOS-Web brown-dwarf candidate catalog of \citep{Festin2026}, and inspect the available COSMOS-3D Unfold products, including both the two-dimensional grism coadds and the associated one-dimensional spectra. Among the candidates with available spectroscopy, COSMOS2025 \footnote{\url{https://cosmos2025.iap.fr/catalog.html}}  ID~639188 was selected for detailed analysis because its F444W trace is centered and relatively isolated, and its local background optimal extraction yields a flux scale consistent with the COSMOS2025 photometry. For readability, we hereafter refer to this source as
\emph{Shabgard}, from the Persian word \textit{shabgard}
(night wanderer), while retaining COSMOS2025 ID~639188 as its formal catalog identifier.

To our knowledge, this is the first spectrophotometric analysis of this source. We extract and validate its F444W WFSS spectrum, combine it with broadband JWST/NIRCam photometry, and compare the data with several brown-dwarf atmosphere and evolutionary model families. We aim to investigate which properties can be constrained by the present data, particularly $\Teff$, surface gravity, metallicity, distance, and mass, and which remain uncertain because of the restricted F444W wavelength coverage.

This paper is organized as follows. In Section~\ref{sec:data}, we describe the photometric candidate selection, the COSMOS2025 catalog measurements, and the COSMOS-3D WFSS data. Section~\ref{sec:extraction} presents the local-background optimal extraction of the F444W spectrum. Section~\ref{sec:methods} describes the atmosphere-model comparison, synthetic photometry, weighting scheme, LOWZ atmosphere-model interpolation, MCMC analysis, and the molecular-opacity and empirical spectral tests. Section~\ref{sec:results} presents the atmosphere-model results, inferred physical parameters, multi-epoch proper-motion measurement, and supplementary machine-learning consistency test. In Section~\ref{sec:discussion}, we discuss the brown-dwarf interpretation, Galactic kinematics, agreement and differences among the model families, and limitations of the current data. We summarize our conclusions in Section~\ref{sec:conclusions}.

\section{Candidate Selection and Data}\label{sec:data}

This section describes how Shabgard (COSMOS2025 ID~639188) was selected and which observational data are used in the analysis. We first summarize the photometric brown-dwarf candidate selection from the COSMOS2025 catalog, then describe the COSMOS-Web photometry and COSMOS-3D NIRCam/F444W WFSS products used for the spectrophotometric analysis. We also explain why Shabgard was chosen for detailed follow-up from the subset of candidates with available grism spectra.

\subsection{COSMOS-Web photometry and COSMOS-3D WFSS spectroscopy}

This work is based on the combination of COSMOS-Web imaging, COSMOS2025 catalog photometry, and COSMOS-3D slitless spectroscopy. The Cosmic Evolution Survey (COSMOS) field is a $\sim2~\mathrm{deg^2}$ equatorial field with extensive multi-wavelength imaging and spectroscopy from X-ray to radio wavelengths \citep{Scoville2007}. This makes it a particularly useful field for identifying rare compact sources and for checking their photometric and morphological properties across many bands.

COSMOS-Web, the JWST Cosmic Origins Survey, is a Cycle~1 Treasury program covering the central COSMOS region with wide-area JWST imaging \citep{Casey2023}. The survey provides NIRCam imaging in F115W, F150W, F277W, and F444W over $\sim0.54~\mathrm{deg^2}$, together with MIRI/F770W imaging over a smaller $\sim0.19~\mathrm{deg^2}$ area \citep{Casey2023,Franco2026}. For the broadband photometric constraints in this paper, we use the COSMOS2025 master catalog \citep{Shuntov2025}. In this paper, the COSMOS2025 photometry is used as the broadband SED constraint, with the F444W point providing the main flux anchor for comparison with the grism spectrum.

The spectroscopic data come from COSMOS-3D, which adds JWST/NIRCam wide-field slitless spectroscopy (WFSS) in F444W across $\sim0.33~\mathrm{deg^2}$ within the COSMOS-Web region (Kakiichi et al. in prep). The spectroscopic data used in this work are the
COSMOS-3D NIRCam/F444W WFSS products generated with the
\texttt{unfold\_jwst} reduction framework
\citep{Wang2026ASPIREEnvironments,Yang2023ASPIRE,Wang2023ASPIRE}. The available products include
two-dimensional spectral coadds containing the science,
inverse-variance, mask, wavelength, and model-trace information,
together with associated one-dimensional coadd spectra. Since the source is contaminated by nearby bright star spikes, we use
the two-dimensional Unfold product as the input to the
local-background, profile-weighted re-extraction described in
Section~\ref{sec:extraction}. The COSMOS-3D observing program and survey design are 
described by Kakiichi et al. (in preparation).

The NIRCam long-wavelength grisms provide slitless spectra with resolving power $R\sim1600$ near $4\mu\mathrm{m}$ \citep{Greene2017}. In the F444W configuration, the first-order spectra cover approximately $3.9$-$5.0\mu\mathrm{m}$ \footnote{See the JWST User Documentation for NIRCam grisms: \url{https://jwst-docs.stsci.edu/jwst-near-infrared-camera/nircam-instrumentation/nircam-grisms}}. This wavelength range provides the spectroscopic basis for the COSMOS-3D analysis used in this paper.

\subsection{Photometric pre-selection}

The target studied in this paper, Shabgard (COSMOS2025 ID~639188), is drawn from the COSMOS-Web photometric brown-dwarf candidate catalog of \citep{Festin2026}. That catalog starts from the COSMOS2025 v1.0 source list, which contains approximately $8\times10^5$ sources over the full $0.54~\mathrm{deg^2}$ COSMOS-Web area, and applies a staged selection designed to isolate compact sources with T/Y-dwarf-like SEDs. In brief, the \citep{Festin2026} selection uses: (1) an F444W signal-to-noise requirement, with F444W emphasized because the $\sim4.5\mu\mathrm{m}$ window is often the highest-S/N JWST/NIRCam band for cool T/Y dwarfs; (2) catalog quality cuts, including \texttt{warn\_flag=0} and rejection of strongly blended or stellar-contamination-flagged sources; (3) COSMOS2025 point-source morphology/type information; (4) color cuts targeting the blue short-wavelength and red F277W-F444W behavior expected for cool brown dwarfs; and (5) visual inspection of JWST cutouts and public COSMOS-Web SED panels. The filtering sequence reduces the parent catalog from $\sim8\times10^5$ sources to 76 brown-dwarf candidates, including Shabgard (ID~639188) \citep{Festin2026}.

We next inspected the available COSMOS-3D Unfold products for the 76 photometric brown-dwarf candidates from \citep{Festin2026}. In the current reduction, 36 of these candidates have associated COSMOS-3D spectral files. Most of these spectra are not suitable for direct atmospheric fitting, either because the continuum signal is too weak, the trace is affected by overlapping slitless spectra, the source lies close to a wavelength edge, or the local background is difficult to define in crowded regions. After inspecting both the two-dimensional coadds and the default one-dimensional spectra, we identified six candidates with centered and sufficiently isolated F444W traces for possible continuum extraction. The extraction diagnostics for these six sources are presented in
Appendix~\ref{app:other_extractions}. ID~639188 is the brightest F444W source among the 76 candidates in
the parent sample and, among the six candidates with suitable traces,
it also gives the most stable extraction and the clearest agreement
between the extracted F444W flux scale and the COSMOS2025 photometry.
We therefore select it, before any atmosphere-model fitting, as the
best-characterized candidate for this case-study analysis, and adopt
Shabgard for the detailed spectrophotometric analysis in this paper.

For Shabgard, we adopt the COSMOS2025 position of
ID~639188, R.A. $=150.0634341^{\circ}$ and
Decl. $=+2.3552216^{\circ}$. The source has \texttt{warn\_flag}=0 and appears compact in the JWST/NIRCam imaging. In the independent photometric selection of \citep{Festin2026}, the object is also favored by the stellar-template solution over the galaxy-template solution, with $\chi^2_{\rm star}=33.58$ and $\chi^2_{\rm galaxy}=147317$ in their LePhare configuration \citep{Arnouts1999,Ilbert2006}. Both fits use the same photometry,
degrees of freedom, and extinction assumptions, allowing a direct
comparison of the two $\chi^2$ values and strongly disfavoring a
normal galaxy SED. The photometry used in the model fitting is summarized in Table~\ref{tab:photometry}. Shabgard is red from F277W to F444W, with F277W-F444W$\simeq3.40$~mag in the current catalog columns.

\begin{deluxetable}{lcccc}
\tablecaption{COSMOS2025 photometry for Shabgard.\label{tab:photometry}}
\tablehead{\colhead{Band} & \colhead{AB mag} & \colhead{AB mag error} & \colhead{$f_\nu$ [$\mu$Jy]} & \colhead{$\sigma_{f_\nu}$ [$\mu$Jy]}}
\startdata
F115W & 24.196 & 0.011 & 0.7615 & 0.0075 \\
F150W & 24.851 & 0.011 & 0.4167 & 0.0042 \\
F277W & 25.615 & 0.019 & 0.2061 & 0.0037 \\
F444W & 22.215 & 0.003 & 4.7216 & 0.0119 \\
\enddata
\tablecomments{Flux densities and uncertainties are taken from the COSMOS2025 catalog columns. Model fitting uses these values with the additional error floors described in Section~\ref{sec:methods}.}
\end{deluxetable}

\section{Local-background Profile-weighted Extraction}\label{sec:extraction}

For our analysis, we perform a custom one-dimensional
re-extraction of Shabgard from the COSMOS-3D
\texttt{unfold\_jwst} two-dimensional \texttt{coadd2d} product,
rather than adopting the supplied \texttt{coadd1d} spectrum
directly. The upstream Unfold products provide the calibrated
two-dimensional science image, inverse-variance map, pixel mask,
wavelength solution, and model-trace information
\citep{Wang2023ASPIRE}. Our subsequent processing is
limited to the local-background estimation, profile-weighted
extraction, masking, and quality checks described below.

This additional extraction step is motivated by the nature of
slitless spectroscopy, in which all sources in the field are
dispersed and faint continuum spectra can be affected by
overlapping spectral wings, residual structure from neighbouring
traces, and a spatially varying background. Similar challenges are
commonly addressed through source-dependent trace definitions,
contamination modelling, and extraction apertures tied to the
direct-image morphology or spectral trace
\citep{Kummel2009,Brammer2012,Momcheva2016}.

\begin{figure*}[ht]
\centering\includegraphics[width=0.98\textwidth]{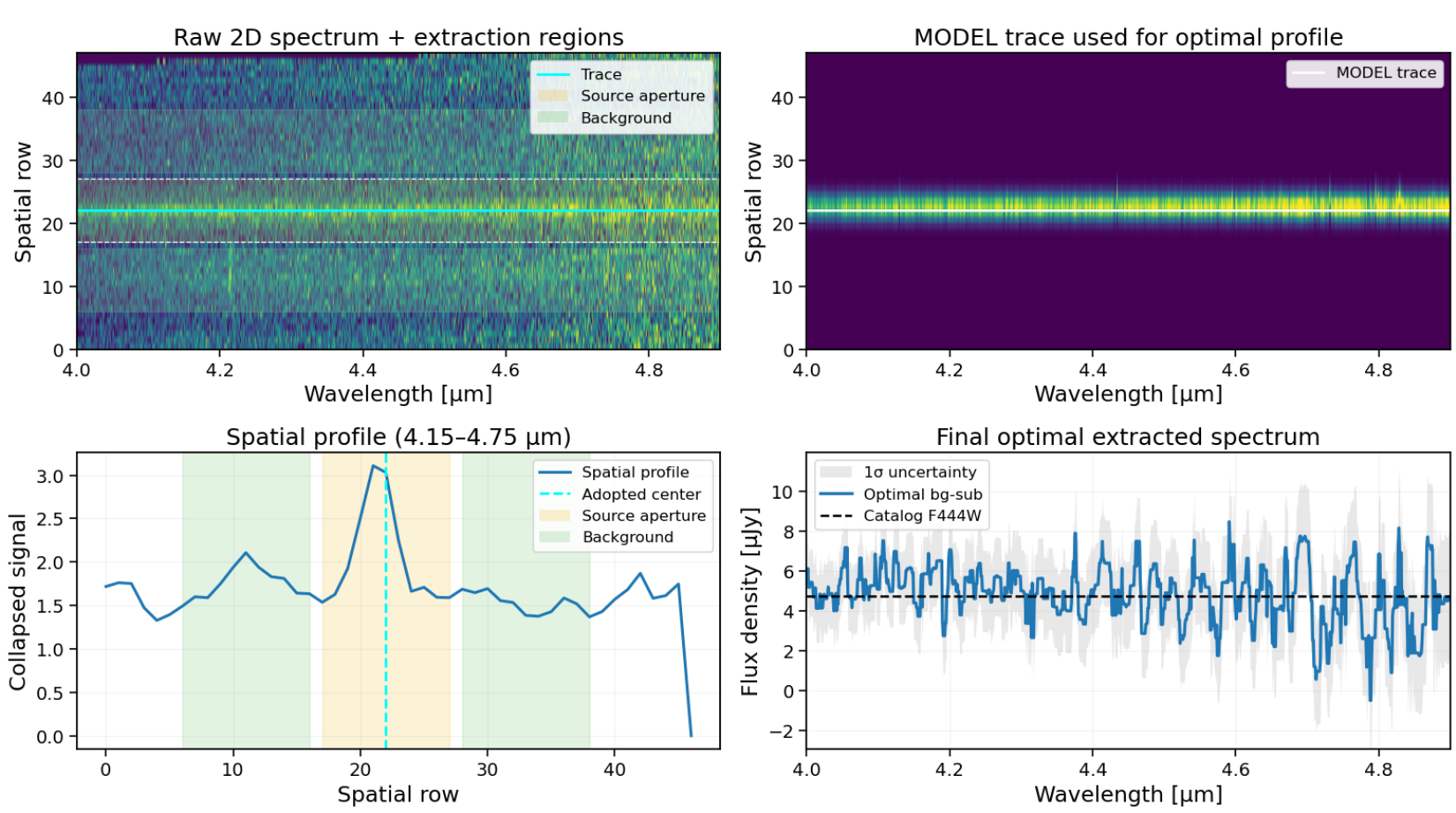}
\caption{Local-background optimal extraction for Shabgard. The upper-left panel shows the two-dimensional F444W WFSS spectrum with the traced source position, source aperture, and local background regions. The upper-right panel shows the corresponding \texttt{MODEL} trace used to define the optimal-extraction profile. The lower-left panel shows the spatial profile collapsed over the central wavelength range, with the adopted source and background regions marked. The lower-right panel shows the final local-background-subtracted optimal one-dimensional spectrum, with the COSMOS2025 F444W catalog flux shown for comparison.}
\label{fig:trace}
\end{figure*}
The extraction geometry for Shabgard is shown in
Figure~\ref{fig:trace}. The initial trace center is determined
automatically from the peak of the wavelength-collapsed
\texttt{MODEL} spatial profile, and the trace is then followed as a
function of wavelength within a $\pm7$-pixel search window and
median-smoothed along the dispersion direction. We adopt a source half-width of 5 spatial pixels. On each side of the wavelength-dependent trace center, candidate background pixels are selected between offsets of 6 and 16 spatial pixels. Thus, the inner background boundary is located approximately one pixel beyond the nominal source-aperture boundary; the value of 6 pixels is measured from the trace center rather than from the edge of the source aperture. Depending on the subpixel trace position, zero or one unused detector row lies between the source and background selections.
These values were chosen after testing reasonable alternatives and were found to produce stable fluxes.

We tested the sensitivity of the extraction to nearby reasonable
geometries by varying the source half-width from 4-6 pixels, the
inner background offset from 5-7 pixels, and the background width from
8-12 pixels. Across the resulting 27 extractions, the synthetic
F444W flux has a median of $4.94~\mu{\rm Jy}$ and a central 68\%
range of $4.78$-$5.09~\mu{\rm Jy}$. The adopted configuration gives
$4.94~\mu{\rm Jy}$, indicating that the extracted flux is not
strongly sensitive to the exact aperture and background geometry.

Before extraction, we construct a valid-pixel mask requiring finite science values, finite positive inverse variance, and a positive \texttt{MASK} value. We also apply a conservative two-dimensional bad-pixel rejection along each spatial row, using a local running median and rejecting only isolated $8\sigma$ outliers. This step is intended to remove hot pixels or narrow detector artifacts without suppressing broad spectral structure.

The local background is estimated independently in each wavelength column. Background pixels are first sigma-clipped using the median and a median-absolute-deviation estimate of the scatter. If valid background pixels are available on both sides of the trace, we fit a weighted linear function of spatial row and evaluate this model across the source aperture. If only one side of the background region is usable, or if the linear fit is unstable, we instead subtract a constant background equal to the sigma-clipped median of the remaining valid background pixels in that column. This provides a local column-by-column background subtraction before the one-dimensional extraction.

The final one-dimensional spectrum is extracted using a profile-weighted estimator following \citep{Horne1986}. For each wavelength column, the spatial profile $P_{ij}$ is taken from the non-negative \texttt{MODEL} image and normalized over the valid source pixels. 
If the model profile is unavailable or pathologically narrow, we use a Gaussian fallback profile centered on the trace with $\sigma=1.4$ spatial pixels. The extracted flux density in column $j$ is then computed as
\begin{equation}
f_j =
\frac{\sum_i P_{ij}[D_{ij}-B_{ij}]w_{ij}}
{\sum_i P_{ij}^{2}w_{ij}},
\end{equation}
where $D_{ij}$ is the two-dimensional science image, $B_{ij}$ is the fitted local background evaluated over the source aperture, and $w_{ij}$ is the inverse-variance weight after masking. The corresponding formal variance is
\begin{equation}
\sigma^2_{f,j} =
\frac{1}{\sum_i P_{ij}^{2}w_{ij}} .
\end{equation}
This formal variance assumes independent pixels and does not account for covariance introduced by resampling of the \texttt{coadd2d} data. For the final fitting vector, we keep only columns with finite wavelength, flux, and uncertainty, at least two valid source pixels, at least four valid background pixels, and wavelengths in the range $4.05$-$4.85~\mu\mathrm{m}$. For Shabgard, this gives 816 usable wavelength pixels out of 1302 columns in the \texttt{coadd2d} spectrum.

As a flux-scale validation, we compare synthetic F444W photometry measured from the extracted spectra with the COSMOS2025 catalog value. The synthetic photometry is computed with the F444W throughput curve using the photon-weighted mean flux density.

The local-background profile-weighted extraction gives synthetic F444W fluxes of $4.858\mu Jy$ over the full usable wavelength range and $4.950\mu Jy$ over the trimmed fitting range, in good agreement with the COSMOS2025 catalog value of $4.7216\mu Jy$. As an independent check, a simple local-background aperture extraction gives similar values of $4.950\mu Jy$ and $4.898\mu Jy$ over the same two wavelength ranges. In contrast, the default pipeline one-dimensional spectrum gives substantially larger synthetic F444W fluxes of $16.31\mu Jy$ and $13.48\mu Jy$. We therefore adopt the local-background profile-weighted spectrum for the atmosphere-model fitting and use the pipeline spectrum only as an extraction diagnostic.

The same extraction diagnostics for the other centered-trace candidates are shown in Appendix Figure~\ref{fig:other_extractions}.

\section{Atmosphere-model Fitting Methods}\label{sec:methods}

The atmosphere-model comparison is designed as a cross-model consistency test rather
than as a formal retrieval from a single atmosphere grid. Because the available
spectrum covers only the NIRCam/F444W grism window, different combinations of
temperature, gravity, metallicity, vertical mixing, and cloud opacity can produce
similar broad-band behavior. We therefore compare Shabgard with six brown-dwarf atmosphere model families and focus on the physical conclusions that
remain stable across different model assumptions.

\subsection{Model families}

The final model comparison includes six atmosphere model families:
Sonora Elf Owl, LOWZ, SAND, Sonora Bobcat, Morley cloudy models, and
ATMO2020/ATMO2020++. These grids were chosen because they test complementary
aspects of cool brown-dwarf atmospheres, including metallicity, surface gravity,
C/O ratio, vertical mixing, cloud-free chemistry, evolutionary consistency, and
possible condensate-cloud opacity. Table~\ref{tab:modelgrid} gives the compact
names, full model names, main physical content, and role of each model family in
this paper.

\begin{table*}[t]
\centering
\caption{Brown-dwarf atmosphere model families used in the final comparison.
\label{tab:modelgrid}}
\scriptsize
\begin{tabular}{llll}
\toprule
Compact name & Full name and reference & Main model physics / parameters & Role in this paper \\
\midrule

\parbox[t]{0.12\textwidth}{Sonora Elf Owl} &
\parbox[t]{0.23\textwidth}{Sonora Substellar Atmosphere Models IV: Elf Owl \citep{Mukherjee2024}} &
\parbox[t]{0.29\textwidth}{Cloud-free, self-consistent atmospheres including disequilibrium chemistry, metallicity, C/O ratio, and vertical mixing parameterized by $K_{zz}$.} &
\parbox[t]{0.23\textwidth}{Tests chemical degeneracies in the 3-5~$\mu\mathrm{m}$ region, especially among metallicity, C/O, and vertical mixing.} \\

\parbox[t]{0.12\textwidth}{LOWZ} &
\parbox[t]{0.23\textwidth}{Low-metallicity ultracool-dwarf atmosphere grid distributed through \texttt{ucdmcmc} \citep{Meisner2021LOWZ,LueberBurgasser2025}} &
\parbox[t]{0.29\textwidth}{Low-temperature and low-metallicity atmosphere models motivated by T-subdwarf studies. The fitted parameters include $T_{\rm eff}$, $\log g$, [M/H], and a dilution factor.} &
\parbox[t]{0.23\textwidth}{Primary low-metallicity test and the model family used for the PCA-interpolated MCMC analysis.} \\

\parbox[t]{0.12\textwidth}{SAND} &
\parbox[t]{0.23\textwidth}{Spectral ANalog of Dwarfs model family \citep{Alvarado2024,LueberBurgasser2025}} &
\parbox[t]{0.29\textwidth}{Ultracool-dwarf atmosphere grid with leverage on metallicity and abundance patterns, including $T_{\rm eff}$, $\log g$, [M/H], and [$\alpha$/Fe].} &
\parbox[t]{0.23\textwidth}{Independent test of a cool, metal-poor brown-dwarf interpretation.} \\

\parbox[t]{0.12\textwidth}{Sonora Bobcat} &
\parbox[t]{0.23\textwidth}{The Sonora Brown Dwarf Atmosphere and Evolution Models. I: cloudless atmospheres in rainout chemical equilibrium \citep{Marley2021}} &
\parbox[t]{0.29\textwidth}{Cloud-free atmosphere and evolutionary models with rainout chemistry, spanning substellar temperatures, gravities, metallicities, radii, masses, and ages.} &
\parbox[t]{0.23\textwidth}{Independent cloud-free atmosphere/evolution check on the preferred temperature and physical scale.} \\

\parbox[t]{0.12\textwidth}{Morley cloudy models} &
\parbox[t]{0.23\textwidth}{Cloudy T/Y-dwarf atmosphere models including sulfide and salt condensate clouds \citep{Morley2012}} &
\parbox[t]{0.29\textwidth}{Cool brown-dwarf atmospheres with condensate-cloud opacity, commonly parameterized by the sedimentation efficiency $f_{\rm sed}$.} &
\parbox[t]{0.23\textwidth}{Tests whether possible cloud opacity can also reproduce the observed SED and F444W continuum shape. This is used as a cloud-systematics check, not as a direct cloud detection.} \\

\parbox[t]{0.12\textwidth}{ATMO2020/++} &
\parbox[t]{0.23\textwidth}{ATMO cool brown-dwarf and giant-exoplanet atmosphere/evolution models \citep{Phillips2020,Leggett2021,Meisner2023}} &
\parbox[t]{0.29\textwidth}{Cool T/Y atmosphere and evolutionary models with an independent treatment of temperature-pressure structure, chemistry, and substellar evolution.} &
\parbox[t]{0.23\textwidth}{Independent model-family check on the preferred temperature, gravity, and evolutionary interpretation.} \\

\bottomrule
\end{tabular}
\end{table*}

The use of multiple grids is important because each family encodes different
assumptions about atmospheric chemistry and opacity. Sonora Elf Owl and Sonora
Bobcat provide cloud-free comparisons, but with different chemical assumptions
and parameter coverage. LOWZ and SAND are especially useful for testing
subsolar-metallicity solutions. The Morley cloudy grid is included because cool
T/Y-dwarf atmospheres can be affected by sulfide and salt condensate clouds;
therefore, we also test whether cloud opacity could plausibly contribute to the
observed spectral shape. Because the current data cover only the F444W grism
interval, we do not interpret the cloud-model comparison as evidence for a
detected cloud feature. Instead, it is used to estimate how strongly the
atmospheric interpretation depends on the choice of cloud-free versus cloudy
models.

\subsection{Synthetic photometry, weighting, and fit metrics}
\label{fit_metrics}
All model spectra are converted to the same observational space before fitting.
Each model is placed on an $F_{\nu}$ flux-density scale, smoothed to the
approximate NIRCam/WFSS resolution, interpolated onto the observed wavelength
grid, and integrated through the JWST/NIRCam filter curves to produce synthetic
photometry. For both the observed spectrum and the models, synthetic band fluxes
are computed using the photon-weighted mean flux density,
\begin{equation}
\langle f_{\nu}\rangle =
\frac{\int f_{\nu}(\lambda)\,T(\lambda)\,\lambda^{-1}\,d\lambda}
{\int T(\lambda)\,\lambda^{-1}\,d\lambda},
\end{equation}
where $T(\lambda)$ is the total system throughput of the filter. This convention
is used consistently for the COSMOS2025 photometry comparison and for synthetic
F444W fluxes measured from the extracted WFSS spectrum.

For each model, we solve for a positive multiplicative scale factor between the
model and the data. For observed fluxes $f_i$, model fluxes $m_i$, uncertainties
$\sigma_i$, and optional relative weights $w_i$, the weighted least-squares
normalization is
\begin{equation}
a =
\frac{\sum_i w_i f_i m_i/\sigma_i^2}
{\sum_i w_i m_i^2/\sigma_i^2}.
\end{equation}
For most model-family comparisons, this scale is treated only as the best
normalization needed to compare model shape and photometry. For surface-flux
model grids, however, the normalization can have a physical interpretation. In
the LOWZ MCMC analysis, we therefore fit the dilution factor
$\alpha=(R/d)^2$, where $R$ is the object radius and $d$ is the distance.

The likelihood uses the local-background, profile-weighted WFSS spectrum over
$4.05$-$4.85~\mu\mathrm{m}$ together with the available JWST/NIRCam photometry. We
adopt empirical error floors of 8\% plus $0.05~\mu{\rm Jy}$ for the spectral
points and 5\% for the photometric points. The spectral floor is conservative compared with
the scatter among the alternative extraction geometries described
above. We also repeated the model comparison using photometric floors
of 3\%, 5\%, and 10\%. The preferred parameters
remain unchanged across this range. These floors are intentionally
conservative because the dominant uncertainty is not purely photon noise:
background subtraction, trace placement, residual slitless contamination, and
model imperfections can all contribute additional systematic residuals.

The F444W photometric point is given the largest relative weight in the
photometric part of the fit because it anchors the absolute flux scale over the
same wavelength region sampled by the WFSS spectrum. This prevents the noisy
grism continuum from moving the model normalization away from the COSMOS2025
broad-band flux, while still allowing the F444W spectral shape to constrain the
atmosphere. The remaining NIRCam bands provide the broader SED context.

For reporting and ranking the grid-search fits, we compute separate mean
spectroscopic and weighted-mean photometric squared-residual diagnostics. The
spectroscopic contribution is defined as
\begin{equation}
\overline{\chi^2}_{\rm spec} =
\frac{1}{N_{\rm spec}}
\sum_{j=1}^{N_{\rm spec}}
\left[
\frac{f_j-a m_j}{\sigma_j}
\right]^2 ,
\end{equation}
where $f_j$ and $m_j$ are the observed and model spectral flux densities,
$\sigma_j$ is the corresponding uncertainty, and $a$ is the fitted
multiplicative normalization.

The photometric contribution is the weighted mean
\begin{equation}
\overline{\chi^2}_{{\rm phot},w} =
\frac{
\sum_b w_b
\left[
\frac{F_b-a M_b}{\sigma_b}
\right]^2
}{\sum_b w_b},
\end{equation}
where $F_b$ and $M_b$ are the observed and synthetic model flux densities
in band $b$, and $w_b$ is the adopted relative band weight. We use
$w_b=1$ for F115W, F150W, and F277W and $w_b=10$ for F444W, so that
F444W provides the strongest photometric constraint over the wavelength
region sampled directly by the WFSS spectrum.

The balanced ranking objective is
\begin{equation}
Q_{\rm bal} =
\overline{\chi^2}_{\rm spec}
+2\overline{\chi^2}_{{\rm phot},w}.
\end{equation}
Because the F444W photometric constraint overlaps the WFSS wavelength
range, and because the relative weights and error floors are empirical,
$Q_{\rm bal}$ is a weighted pseudo-$\chi^2$ objective used only to rank
models consistently. It is not interpreted using a formal
$\chi^2$ probability distribution. 

The multiplicative normalization $a$ is solved by minimizing this same
objective, so the adopted spectral, photometric, and band weights are applied
consistently in both the normalization and the model ranking. The resulting
$Q_{\rm bal}$ values are used for relative comparison of the discrete
atmosphere-grid solutions; quantitative parameter uncertainties for LOWZ are
derived separately from the likelihood-based MCMC analysis. The fitted vector contains 816 spectral points and four photometric
points. For each fixed grid model, one multiplicative normalization is
fitted, corresponding nominally to 819 residual degrees of freedom.
However, spectral resampling covariance and the overlapping F444W
constraints prevent these from being interpreted as independent formal
degrees of freedom.

In addition to $\chi^2$, we compute the cosine similarity of the WFSS spectral
shape,
\begin{equation}
\cos\theta =
\frac{\sum_j f_j m_j}
{\left(\sum_j f_j^2\right)^{1/2}
 \left(\sum_j m_j^2\right)^{1/2}} .
\end{equation}
Cosine similarity is a standard vector-space similarity measure originally
developed in information retrieval and pattern matching \citep{Salton1975}.
Here it is used only as a secondary diagnostic of spectral-shape agreement. The
model ranking and physical interpretation are not based on cosine similarity
alone.

\subsection{LOWZ PCA-interpolated MCMC}

The LOWZ atmosphere grid is discrete, so we constructed a continuous
interpolator before sampling. We fixed C/O$=0.55$ and
$\log K_{zz}=-1.0$ to the values of the preferred discrete LOWZ
solution and interpolated over $T_{\rm eff}$, $\log g$, and [M/H].
After fixing these chemistry parameters, the grid contains 17 values of
$T_{\rm eff}$ spanning 500-1600~K, five values of $\log g$ spanning
3.5-5.25, and 11 values of [M/H] spanning $-2.5$ to $+1.0$.
Of the 935 possible grid nodes, 931 are available, corresponding to
99.6\% completeness. The four missing nodes were filled by local
interpolation from the surrounding available models before constructing
the regular interpolator.

At each grid node, we formed the same 820-element model vector used in
the likelihood, containing the 816 F444W spectral elements and the four
NIRCam photometric fluxes. The spectral and photometric components were
scaled to reproduce the relative weighting defined in
Section~4.2. Before applying principal-component analysis
\citep[PCA;][]{JolliffeCadima2016}, each element was divided by its
corresponding uncertainty and the model vectors were mean-centered. We
retained the minimum number of principal components required to explain
99.99\% of the weighted model variance. Six components were required,
retaining 99.9909\% of the variance. The PCA coefficients were then
linearly interpolated in
$(T_{\rm eff},\log g,[{\rm M/H}])$ space, and the full model vector was
reconstructed at each proposed MCMC position. This procedure
interpolates the model spectra and photometry themselves rather than
interpolating fitted parameters or discrete fit statistics.

The sampled parameters were $T_{\rm eff}$, $\log g$, [M/H], and
$\log_{10}\alpha$, where $\alpha=(R/d)^2$. We adopted uniform priors
within the LOWZ grid limits given above and a uniform prior
$-26<\log_{10}\alpha<-16$. The likelihood was Gaussian and used the
same observational error floors and spectral-to-photometric weighting
as the discrete model comparison. To account conservatively for
residual atmosphere-model and interpolation uncertainties, an
additional 3\% systematic uncertainty was added in quadrature to each
element of the weighted data-vector uncertainty. This floor was
calculated relative to the larger of the absolute weighted datum and
the median absolute value of the weighted data vector.

We sampled the posterior using \texttt{emcee}
\citep{ForemanMackey2013} with 80 walkers. The walkers were initialized
near the preferred discrete LOWZ solution using a random seed of 42.
We discarded an initial 1200-step burn-in phase and then ran each walker
for 3000 production steps, thinning the resulting chains by a factor
of 10. The estimated integrated autocorrelation times were
approximately 45-47 steps for the four sampled parameters, so the
production chains extended over more than 60 autocorrelation times.

C/O and $K_{zz}$ were fixed rather than fitted simultaneously because
the restricted F444W wavelength coverage does not provide sufficient
independent leverage to robustly separate metallicity, C/O ratio,
vertical mixing, surface gravity, and flux normalization. The resulting
posterior is therefore interpreted as a LOWZ-based physical-scale
estimate under fixed-chemistry assumptions, rather than as a complete
atmospheric retrieval.

\subsection{Molecular-opacity and empirical spectral tests}
\label{sec:molecular_tests}

To examine the molecular origin of the structure in the F444W spectrum,
we carried out two complementary tests. First, we use representative
ExoMolOP opacity profiles \citep{Chubb2021} for CH$_4$, CO$_2$, and CO at
$T=700$~K and $P=1$~bar as qualitative wavelength guides. Because
cool-brown-dwarf molecular bands contain dense and overlapping
rovibrational transitions, the opacity profiles are displayed at
$R\sim300$ to suppress individual line structure and emphasize the broad
wavelength dependence of each opacity source. Each opacity curve is
independently normalized, inverted, and rescaled to the Shabgard flux
range for visualization only. No opacity-to-flux conversion is implied,
and the observed Shabgard spectrum remains at its native
NIRCam/F444W WFSS resolving power of $R\sim1600$.

We then performed a fixed-atmosphere forward-model test using PICASO \citep{Batalha2019} and
the pressure-temperature and native chemistry profile of the
best-ranked Sonora Elf Owl V2 \citep{Mukherjee2024,Wogan2025} grid solution
($T_{\rm eff}=700$~K, $\log g=5.50$, $[\mathrm{M/H}]=-0.50$,
C/O$=1.0$, and $\log K_{zz}=2.0$).
The native composition is H$_2$/He dominated. For the molecular-opacity and empirical spectral comparisons only,
we use the $R\sim1600$ binned representation of the extracted
spectrum, containing 289 samples over 4.05-4.85~$\mu$m.
The original 816-pixel extraction remains the input to the main
spectro-photometric atmosphere-model fitting described in
Section~\ref{fit_metrics}. The PICASO thermal
spectrum was calculated using the high-resolution $R=15000$ opacity
database, convolved to $R\sim1600$, and sampled on the same
289-point, 4.05-4.85~$\mu$m wavelength grid as Shabgard.
Only one positive multiplicative normalization was fitted; we applied
no wavelength shift, continuum tilt, feature-specific masking, or local
rescaling.

To test molecular attribution, the calculation was repeated after
removing CH$_4$, CO$_2$, and CO opacity individually while keeping the
pressure-temperature structure and the remaining chemistry fixed.
We quantify the change using
\begin{equation}
    \Delta\chi^2 =
\chi^2_{\rm raw,no\,mol}-\chi^2_{\rm raw,full},
\end{equation}
where $\chi^2_{\rm raw}$ is the summed uncertainty-weighted squared
residual. The significance of each change was calibrated with
5000 parametric-bootstrap realizations under the corresponding
no-molecule model. We additionally repeated the comparison for matched
resolving powers of $R=1400$, 1600, and 1800.

As an empirical comparison independent of the atmosphere-model family,
we also compare Shabgard with the published JWST spectra of the
T8 dwarf 2MASS~J0415$-$0935
\citep{AlejandroMerchan2025}
and the T6 dwarf JADES-GS-BD-11
\citep{Hainline2026}.
The 2MASS~J0415$-$0935 spectrum was obtained with
JWST/NIRSpec G395H at $R\sim2700$
\citep{AlejandroMerchan2025,Jakobsen2022NIRSpec}
and is degraded to the native Shabgard information scale
($R\sim1600$).
JADES-GS-BD-11 was observed with NIRSpec/PRISM
\citep{Hainline2026}; for this comparison we degrade Shabgard
to a representative PRISM information scale of $R\sim100$,
noting that the instrumental PRISM resolving power is
wavelength dependent \citep{Jakobsen2022NIRSpec}.
In each comparison we fit only one positive multiplicative
normalization and calculate the diagnostics on the unsmoothed,
resolution-matched vectors.

\section{Results}\label{sec:results}

\subsection{Cross-model atmosphere comparison}

\begin{figure*}[t]
\centering
\includegraphics[width=\textwidth]{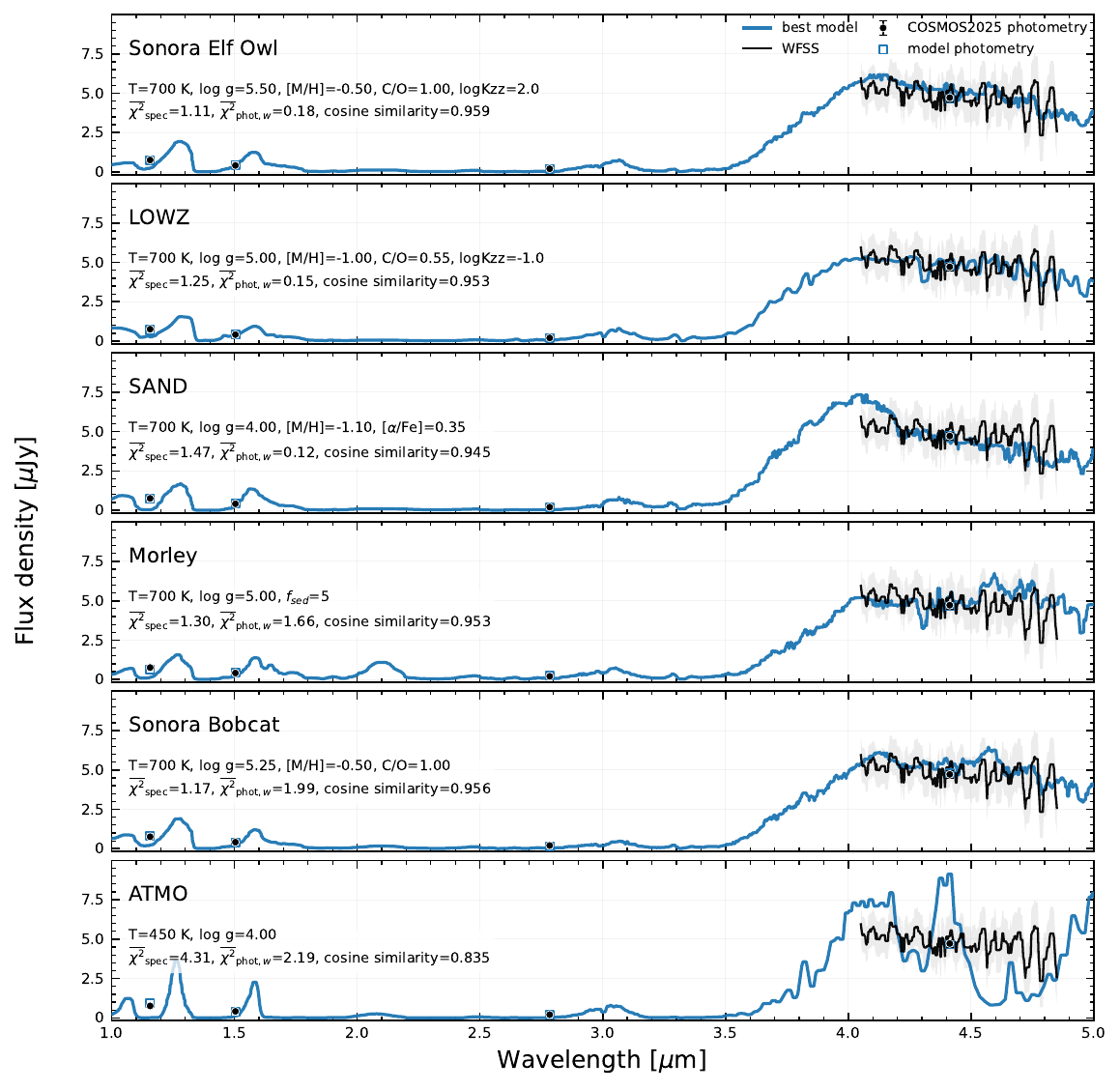}
\caption{
Best-fitting atmosphere models for Shabgard from the six model
families used in the final comparison. The panels are ordered by the adopted
balanced fit score, which combines the WFSS spectral agreement and the
JWST/NIRCam photometric agreement. In each panel, the blue curve shows the
scaled best-fitting model spectrum, the black curve shows the
local-background, profile-weighted F444W WFSS spectrum, and the gray region
shows the spectral uncertainty. Black points show the COSMOS2025 photometry,
and open blue squares show the corresponding synthetic model photometry. The
model families and references are listed in Table~\ref{tab:modelgrid}. All six
model families remain within the cool brown-dwarf regime. Sonora Elf Owl,
LOWZ, SAND, Sonora Bobcat, and Morley converge on
$T_{\rm eff}=700$~K, while ATMO provides a cooler
$T_{\rm eff}=450$~K solution. The Morley cloudy model provides a useful test
of possible condensate-cloud opacity.
}
\label{fig:six_model_comparison}
\end{figure*}

Figure~\ref{fig:six_model_comparison} summarizes the final atmosphere-model
comparison for Shabgard. The strongest result is the convergence of five
independently constructed model families on $T_{\rm eff}=700$~K: Sonora Elf
Owl, LOWZ, SAND, Morley, and Sonora Bobcat. These grids treat chemistry,
metallicity, vertical mixing, clouds, and evolutionary constraints differently,
so their agreement establishes approximately $700$~K as the principal
cross-model temperature scale. ATMO selects a cooler $T_{\rm eff}=450$~K
solution, providing a model-family-dependent lower-temperature comparison while
remaining within the cool brown-dwarf regime.

The surface gravities are moderate to high, although the exact value remains
model-dependent. The displayed solutions span $\log g\sim4.0$-$5.5$:
Sonora Elf Owl gives $\log g=5.50$, Sonora Bobcat gives 5.25, LOWZ and
Morley give 5.00, and SAND and ATMO give 4.00. Thus, most of the highest-ranked
solutions favor moderate-to-high gravity, and none requires an extremely
low-gravity, inflated atmosphere. This is consistent with the interpretation of
Shabgard as a field brown dwarf rather than a very young object.

The metallicity-sensitive grids favor subsolar metallicity, although the exact
abundance is model-dependent. LOWZ gives a displayed best fit with
[M/H]$=-1.00$, and SAND gives [M/H]$=-1.10$. Sonora Elf Owl and Sonora
Bobcat also select subsolar metallicity, with [M/H]$=-0.50$ in both displayed
best fits. In the Sonora grids used here, however, only three metallicity values
are available, and [M/H]$=-0.50$ is the lowest value in the fitted grid. The
Sonora fits should therefore be interpreted as showing a preference toward the
low-metallicity edge of the available Sonora parameter space, rather than as a
precise measurement of [M/H]$=-0.50$. Taken together with the lower-metallicity
LOWZ and SAND solutions, this behavior supports a metal-poor
interpretation, but the current F444W-only spectrum cannot fully separate
metallicity from C/O ratio, surface gravity, and vertical mixing.

The Morley cloudy model gives a fit quality comparable to the cloud-free Sonora
Bobcat model. This does not mean that a cloud feature is detected. Rather, it
shows that a cloudy cool-dwarf atmosphere is allowed by the present data. In
the Morley/Ackerman-Marley cloud framework, the sedimentation-efficiency
parameter $f_{\rm sed}$ controls the vertical thickness of the cloud:
smaller $f_{\rm sed}$ values correspond to less efficient settling and thicker,
more vertically extended clouds, while larger $f_{\rm sed}$ values correspond
to more efficient settling and relatively thinner clouds
\citep{AckermanMarley2001,Morley2012}. The best displayed Morley model has
$f_{\rm sed}=5$, so it should be interpreted as a cloudy but relatively
settled/thinner-cloud solution within that model grid. The F444W-only spectrum
does not provide enough wavelength leverage to distinguish this uniquely from
cloud-free solutions.

\begin{table*}[ht!]
\centering
\caption{
Best-fit atmosphere-model comparison and physical-scale estimates for
Shabgard. The first six rows correspond to the six displayed models in
Figure~\ref{fig:six_model_comparison}, ordered by the adopted balanced fit
score. 
\label{tab:results}}
\scriptsize
\setlength{\tabcolsep}{2.5pt}
\renewcommand{\arraystretch}{1.18}
\begin{tabular}{lllllllll}
\toprule
Model &
$\Teff$ [K] &
$\logg$ &
$\mh$ &
Other parameters &
Physical scale / interpretation &
$\overline{\chi^2}_{\rm spec}$ &
$\overline{\chi^2}_{{\rm phot},w}$ &
Cosine \\
\midrule

\parbox[t]{0.095\textwidth}{Sonora Elf Owl} &
700 &
5.50 &
$-0.50$ &
\parbox[t]{0.14\textwidth}{C/O$=1.00$, $\log K_{zz}=2.0$} &
\parbox[t]{0.18\textwidth}{Best balanced score; cloud-free disequilibrium-chemistry solution.} &
1.11 &
0.18 &
0.959 \\

\parbox[t]{0.095\textwidth}{LOWZ grid} &
700 &
5.00 &
$-1.00$ &
\parbox[t]{0.14\textwidth}{C/O$=0.55$, $\log K_{zz}=-1.0$} &
\parbox[t]{0.18\textwidth}{Low-metallicity grid solution; used as the basis for the MCMC posterior below.} &
1.25 &
0.15 &
0.953 \\

\parbox[t]{0.095\textwidth}{SAND} &
700 &
4.00 &
$-1.10$ &
\parbox[t]{0.14\textwidth}{$[\alpha/{\rm Fe}]=0.35$} &
\parbox[t]{0.18\textwidth}{Independent metal-poor solution; strongest photometric agreement among the displayed models.} &
1.47 &
0.12 &
0.945 \\

\parbox[t]{0.095\textwidth}{Morley cloudy} &
700 &
5.00 &
\nodata &
\parbox[t]{0.14\textwidth}{$f_{\rm sed}=5$} &
\parbox[t]{0.18\textwidth}{Cloudy but relatively settled/thinner-cloud solution; cloud opacity allowed but not uniquely detected.} &
1.30 &
1.66 &
0.953 \\

\parbox[t]{0.095\textwidth}{Sonora Bobcat} &
700 &
5.25 &
$-0.50$ &
\parbox[t]{0.14\textwidth}{C/O$=1.00$} &
\parbox[t]{0.18\textwidth}{Cloud-free atmosphere comparison; a separate evolutionary-scale check is given in Appendix~\ref{app:evolution_checks}.} &
1.17 &
1.99 &
0.956 \\

\parbox[t]{0.095\textwidth}{ATMO} &
450 &
4.00 &
\nodata &
\parbox[t]{0.14\textwidth}{Displayed grid fit} &
\parbox[t]{0.18\textwidth}{Cooler atmosphere-grid solution; a separate evolutionary-scale check is given in Appendix~\ref{app:evolution_checks}.} &
4.31 &
2.19 &
0.835 \\

\parbox[t]{0.095\textwidth}{LOWZ MCMC} &
\parbox[t]{0.08\textwidth}{$718.6^{+10.6}_{-10.2}$} &
\parbox[t]{0.08\textwidth}{$5.13^{+0.09}_{-0.15}$} &
\parbox[t]{0.08\textwidth}{$-1.00^{+0.07}_{-0.08}$} &
\parbox[t]{0.14\textwidth}{$\log_{10}\alpha=-21.76^{+0.02}_{-0.02}$} &
\parbox[t]{0.18\textwidth}{With $R=1.00\pm0.10\rjup$: $d=176^{+18}_{-18}$~pc and $M=52^{+18}_{-16}\mjup$.} &
\nodata &
\nodata &
\nodata \\

\bottomrule
\end{tabular}
\end{table*}

Table~\ref{tab:results} gives the numerical comparison of the best-fit
atmosphere models. Sonora Elf Owl is the highest-ranked displayed model and
combines the lowest balanced score with the highest cosine similarity. LOWZ and
SAND follow closely and independently reinforce the preference for subsolar
metallicity. Morley and Sonora Bobcat provide similarly strong spectral-shape
agreement, showing that the present data allow both a cloudy-but-settled and a
cloud-free atmosphere at approximately $700$~K. ATMO provides a cooler
$450$~K comparison solution that extends the model-family temperature range;
the convergence of the other five families at $700$~K remains the central
cross-model result.

\subsection{LOWZ posterior and physical scale}
\label{lowz_mcmc}
\begin{figure*}[ht!]
\centering
\includegraphics[width=0.92\textwidth]{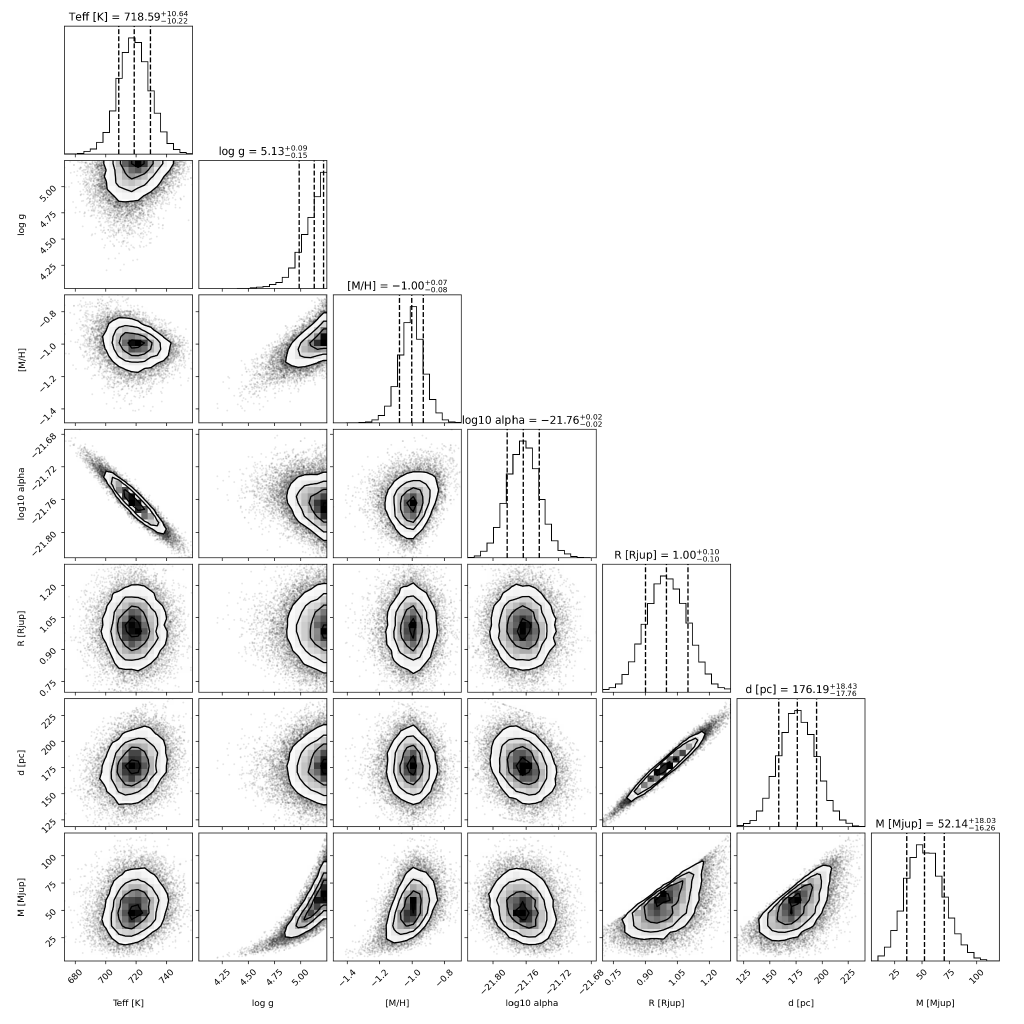}
\caption{
LOWZ PCA-interpolated posterior for Shabgard, including derived physical
parameters. The sampled atmosphere parameters are $T_{\rm eff}$, $\log g$,
[M/H], and $\log_{10}\alpha$, where $\alpha=(R/d)^2$. The radius is not fitted
directly by the LOWZ atmosphere grid; it is drawn from an external
$R=1.00\pm0.10\rjup$ prior. The distance follows from the dilution factor
$\alpha=(R/d)^2$, and the mass follows from $g=GM/R^2$. These physical values
are therefore conditional on both the LOWZ atmosphere solution and the adopted
radius prior.
}
\label{fig:lowzphys}
\end{figure*}

The LOWZ PCA-interpolated MCMC provides the only continuous posterior used in this analysis. The resulting posterior distributions and derived physical quantities are shown in Figure~\ref{fig:lowzphys}. In this fit, the sampled parameters are $T_{\rm eff}$, $\log g$, [M/H], and the dilution factor $\alpha$, while C/O and $K_{zz}$ are fixed to the values of the best LOWZ grid solution. This fixed-chemistry setup is used because the F444W-only spectrum does not provide enough independent wavelength leverage to separate metallicity, C/O ratio, vertical mixing, gravity, and normalization simultaneously. 

To validate the interpolation, we performed leave-one-interior-slice-out
cross-validation in $T_{\rm eff}$, $\log g$, and [M/H], rebuilding the PCA
basis and interpolator after removing each parameter slice. Six principal
components retain 99.9909\% of the model variance. Across 2141 withheld LOWZ
models, the F444W spectral interpolation has an RMS error of
$0.175\,\sigma_{\rm obs}$; near the inferred posterior, the median RMS decreases
to $0.077\,\sigma_{\rm obs}$, with a 95th percentile of
$0.094\,\sigma_{\rm obs}$. 

The LOWZ posterior gives
\begin{align}
\Teff &= 718.6^{+10.6}_{-10.2}\,\mathrm{K},\\
\logg &= 5.13^{+0.09}_{-0.15},\\
\mh &= -1.00^{+0.07}_{-0.08},\\
\log_{10}\alpha &= -21.76^{+0.02}_{-0.02}.
\end{align}
Using a radius prior of $R=1.00\pm0.10\rjup$ \citep{Baraffe2003,SaumonMarley2008} to convert
$\alpha=(R/d)^2$ gives
\begin{align}
d &= 176.2^{+18.4}_{-17.8}\ {\rm pc},\\
M &= 52.1^{+18.0}_{-16.3}\mjup,
\end{align}
where the mass is computed from $g=GM/R^2$. 

The LOWZ physical scale is higher in mass and distance than the ATMO/Sonora
evolutionary checks. ATMO and Sonora Bobcat provide distances near $143$-$145$ pc
and masses near $23\mjup$, while LOWZ gives $d\sim176$ pc and
$M\simeq52\mjup$. This difference is not surprising. LOWZ is used here as an
atmosphere-only low-metallicity grid with a radius prior, whereas ATMO and
Sonora Bobcat provide internally linked atmosphere/evolution solutions. The
two approaches, therefore, respond differently to gravity, normalization, radius,
and metallicity. Importantly, both physical scales remain substellar and place
Shabgard as a foreground Galactic object along the COSMOS line of sight.

\subsection{Multi-epoch proper motion and tangential velocity of
Shabgard}
\label{sec:proper_motion}

The availability of multi-epoch JWST/NIRCam imaging provides an
astrometric test of whether Shabgard is a foreground Galactic source.
We measured its position in PRIMER
\citep{Donnan2024PRIMER}, COSMOS-Web
\citep{Casey2023,Franco2026}, and COSMOS-3D
\citep{Meyer2026COSMOS3D} imaging obtained between
2023 and 2025. The primary measurements use the same pair of observing
epochs, PRIMER and COSMOS-3D, but are obtained independently from the
F115W, F200W, and F356W images. The exposure midpoints recorded in the
original image headers give baselines of approximately $1.927$~yr for
these comparisons. We also use the F115W
COSMOS-Web-COSMOS-3D measurement as a secondary astrometric check.

For each image, the source centroid was measured in a locally
background-subtracted cutout using a two-dimensional Gaussian
estimator. Small residual offsets between epochs were corrected using
nearby compact reference sources. For each pair of images, we
calculated the median displacement of the reference-source population
and subtracted this local-frame offset from the displacement measured
for Shabgard. Following the usual astrometric convention, the
proper-motion components are
\begin{equation}
\mu_{\alpha^\ast}
=
\frac{\Delta\alpha^\ast_{\rm corr}}{\Delta t},
\qquad
\mu_{\delta}
=
\frac{\Delta\delta_{\rm corr}}{\Delta t},
\label{eq:proper_motion_components}
\end{equation}
where
$\Delta\alpha^\ast=\Delta\alpha\cos\delta$ and $\Delta t$ is the
elapsed time between the observations
\citep{Lindegren2012,Butkevich2014}.

\subsubsection{Annual-parallax correction}

The apparent displacement between two epochs can contain contributions
from both the intrinsic proper motion and annual parallax. We therefore
calculated the parallax contribution using the exposure midpoint and
the barycentric JWST coordinates recorded in the
\texttt{MJD-AVG}, \texttt{JWST\_X}, \texttt{JWST\_Y}, and
\texttt{JWST\_Z} keywords of the original \texttt{i2d} products.
Astrometric motion relative to the Solar-system barycentre can be
written in terms of the parallax factors in the local right-ascension
and declination directions \citep{Lindegren2012,Butkevich2014}. For
two epochs, the differential parallax displacement is
\begin{equation}
\Delta\alpha^\ast_{\varpi}
=
\varpi
\left(
P_{\alpha^\ast,2}-P_{\alpha^\ast,1}
\right),
\qquad
\Delta\delta_{\varpi}
=
\varpi
\left(
P_{\delta,2}-P_{\delta,1}
\right),
\label{eq:parallax_displacement}
\end{equation}
where $P_{\alpha^\ast}$ and $P_{\delta}$ are the dimensionless
parallax factors and
\begin{equation}
\varpi\,[{\rm mas}]
=
\frac{1000}{d\,[{\rm pc}]}.
\label{eq:parallax_distance}
\end{equation}
The two parallax terms in
Equation~\ref{eq:parallax_displacement} were subtracted from the
locally registered displacement before calculating the proper motion.

At a representative distance of 160~pc, the predicted differential
parallax displacement for the primary PRIMER-COSMOS-3D epoch pair is
only $0.62$~mas. Over the model-dependent distance range of
approximately 143-176~pc, the correction remains between about
$0.57$ and $0.70$~mas. Its contribution to the inferred proper motion
is therefore only about $0.32\ {\rm mas\,yr^{-1}}$, and the change in
the total proper-motion amplitude is approximately $0.1\%$.

The COSMOS-Web and COSMOS-3D observations sample more widely separated
parallax phases. Their predicted differential parallax displacement is
approximately $10.1$~mas at 160~pc and is consequently more sensitive
to the adopted distance. We therefore use this shorter-baseline
measurement only as a consistency check and do not include it in the
primary combined proper motion.

\begin{figure*}[ht!]
\centering
\includegraphics[width=0.96\textwidth]
{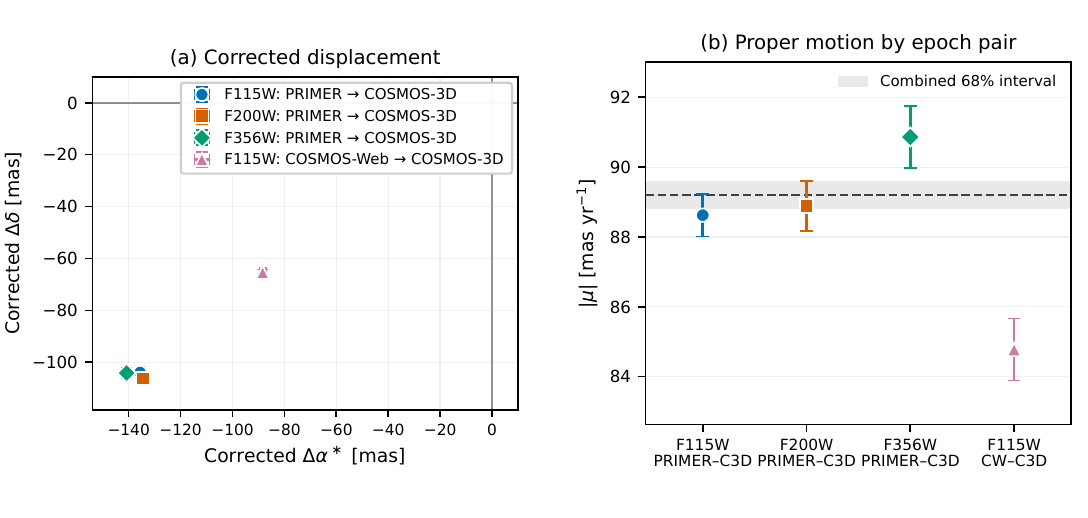}
\caption{
Multi-epoch astrometric measurements for COSMOS2025 Shabgard after
local-frame registration and annual-parallax correction.
Panel~(a) shows the displacement components for the three same-filter
PRIMER-COSMOS-3D comparisons and the additional F115W
COSMOS-Web-COSMOS-3D comparison. Horizontal and vertical error bars
show the robust $1\sigma$-equivalent scatter obtained from the
recovery of injected artificial sources. Panel~(b) shows the
corresponding total proper-motion amplitudes. The dashed horizontal
line and shaded region indicate the median and central 68\% interval
of the combined result from the three PRIMER-COSMOS-3D measurements.
The COSMOS-Web-COSMOS-3D measurement is shown as an independent
check and is not included in the combined value.
}
\label{fig:id639188_proper_motion}
\end{figure*}

\subsubsection{Proper-motion measurement}

The three primary measurements show consistent motion toward
decreasing right ascension and declination, as illustrated in
Figure~\ref{fig:id639188_proper_motion}. Evaluating the small
parallax correction at the representative distance of 160~pc, the
F115W comparison gives
\begin{equation}
(\mu_{\alpha^\ast},\mu_{\delta})
=
(-70.36,-53.89)\ {\rm mas\,yr^{-1}},
\end{equation}
corresponding to
\begin{equation}
|\mu|_{\rm F115W}
=
88.62\pm0.60\ {\rm mas\,yr^{-1}}.
\end{equation}
For F200W, we obtain
\begin{equation}
(\mu_{\alpha^\ast},\mu_{\delta})
=
(-69.76,-55.09)\ {\rm mas\,yr^{-1}},
\end{equation}
with
\begin{equation}
|\mu|_{\rm F200W}
=
88.89\pm0.70\ {\rm mas\,yr^{-1}}.
\end{equation}
The F356W comparison gives
\begin{equation}
(\mu_{\alpha^\ast},\mu_{\delta})
=
(-73.05,-54.03)\ {\rm mas\,yr^{-1}},
\end{equation}
and
\begin{equation}
|\mu|_{\rm F356W}
=
90.86\pm0.90\ {\rm mas\,yr^{-1}}.
\end{equation}

Combining the three PRIMER-COSMOS-3D measurements gives
\begin{equation}
|\mu|
=
89.17^{+0.40}_{-0.40}\ {\rm mas\,yr^{-1}}.
\label{eq:combined_proper_motion}
\end{equation}
The distance dependence of the parallax correction is very small for
this epoch pair. Repeating the calculation with the ATMO, Sonora
Bobcat, and LOWZ distance distributions changes the median combined
proper motion by less than $0.02\ {\rm mas\,yr^{-1}}$.

We assessed the significance of the displacement through
artificial-source injection and recovery. For each epoch pair, 500
sources constructed from an empirical representation of the target
point-spread function were inserted at clean positions with
approximately the same brightness as Shabgard. Two tests were
performed: one in which the artificial sources were displaced by the
measured motion and another in which they remained stationary. The
scatter in the recovered proper motions of the moving artificial
sources is approximately $0.6$-$0.9\ {\rm mas\,yr^{-1}}$ for the
three primary comparisons.

After scaling the recovery distributions to the header-derived time
intervals, the 95th percentiles of the apparent motions of stationary
sources are approximately 3.2, 5.0, and
$27.1\ {\rm mas\,yr^{-1}}$ in F115W, F200W, and F356W,
respectively. The measured amplitudes exceed these limits by factors
of approximately 28, 18, and 3.4. The F115W and F200W measurements
therefore provide particularly strong evidence that the displacement
is not produced by centroid noise or local image structure. The F356W
test is less constraining because its stationary-source distribution
is broader, although its direction and amplitude remain consistent
with the other two filters.

Because the three measurements share the same survey epochs and parts
of the same local astrometric reference frame, the uncertainty in
Equation~\ref{eq:combined_proper_motion} should be interpreted as an
internal injection-recovery uncertainty. Correlated registration
errors could introduce an additional contribution that is not fully
represented by the formal combined interval.

\subsubsection{Tangential velocity and Galactic context}

The measured proper motion was converted to a heliocentric tangential
velocity using
\begin{equation}
V_{\rm tan}
=
4.74047\,
\mu\,d,
\label{eq:tangential_velocity}
\end{equation}
where $\mu$ is in arcsec yr$^{-1}$, $d$ is in pc, and
$V_{\rm tan}$ is in ${\rm km\,s^{-1}}$. This conversion is the
standard relation between angular motion, distance, and transverse
velocity \citep{Johnson1987,Chen2025}.

We propagated the distance and astrometric uncertainties using
$2\times10^{5}$ Monte Carlo realizations. In each realization, we
first drew a distance from the relevant model distribution and
converted it to a parallax. The corresponding parallax displacement
was then calculated separately in right ascension and declination and
subtracted from each measured epoch-pair displacement. The three
corrected proper-motion measurements were combined, and
Equation~\ref{eq:tangential_velocity} was evaluated using the same
distance draw. This procedure retains the weak covariance between
distance, annual parallax, proper motion, and tangential velocity.

The representative ATMO and Sonora Bobcat distances of 143 and 145~pc
give
\begin{equation}
V_{\rm tan,ATMO}
\simeq
60.5\ {\rm km\,s^{-1}}
\end{equation}
and
\begin{equation}
V_{\rm tan,Sonora}
\simeq
61.3\ {\rm km\,s^{-1}},
\end{equation}
respectively. 

For the LOWZ distance posterior,
$d=176.2^{+18.4}_{-17.8}$~pc, the self-consistent calculation gives
\begin{equation}
V_{\rm tan,LOWZ}
=
74.5^{+7.7}_{-7.5}\ {\rm km\,s^{-1}}.
\end{equation}
The model-dependent median range is therefore approximately
\begin{equation}
V_{\rm tan}
\simeq
60\text{-}74\ {\rm km\,s^{-1}}.
\label{eq:velocity_range}
\end{equation}
Recalculating the parallax correction for every distance realization
changes the median tangential velocity by less than
$0.01\ {\rm km\,s^{-1}}$ relative to a calculation using a fixed
160-pc correction. The uncertainty in $V_{\rm tan}$ is therefore
controlled mainly by the model-dependent distance rather than by
annual parallax.

For comparison, the volume-limited ultracool-dwarf sample of
\citet{Schmidt2007} has a mean tangential velocity of
$31.5\ {\rm km\,s^{-1}}$ and a dispersion of
$20.7\ {\rm km\,s^{-1}}$. The velocity inferred for Shabgard is
therefore elevated relative to the mean local ultracool-dwarf
population. It is also substantially higher than the
$12$-$17\ {\rm km\,s^{-1}}$ transverse velocities that
\citet{Chen2025} associated with three low-velocity thin-disk brown
dwarfs in COSMOS-Web. In contrast, a distant galaxy would have an
effectively negligible measurable proper motion over the
approximately 1.93-yr observational baseline, so the large and
consistent motion measured for Shabgard strongly disfavors a distant
extragalactic interpretation.

On the other hand, Shabgard remains below the
$V_{\rm tan}>100\ {\rm km\,s^{-1}}$ high-velocity population
identified by \citet{Faherty2009}, whose kinematics were interpreted
as being associated with older thick-disk or halo ultracool dwarfs.
The present result is therefore most naturally described as
consistent with an older Galactic-disk object and possibly compatible
with thick-disk membership. This interpretation is also qualitatively
consistent with the subsolar-metallicity atmosphere solutions, but
the tangential velocity alone does not provide a unique population
classification.

A radial-velocity measurement is needed to calculate the full
heliocentric $(U,V,W)$ space motion and to perform a quantitative
thin-disk, thick-disk, or halo membership test
\citep{Johnson1987,Faherty2009}. The significant motion detected
independently in three NIRCam filters nevertheless strongly disfavors
a stationary extragalactic interpretation and provides separate
support for identifying Shabgard as a foreground Galactic ultracool
dwarf.
\subsection{Supplementary Random-forest Regression Cross-check}
\label{sec:lowz_rf}

As a supplementary check on the LOWZ-based atmospheric inference, we trained regression models using the fixed-chemistry LOWZ atmosphere grid \citep{Meisner2021LOWZ}, accessed with \texttt{ucdmcmc} v1.4.4 \citep{Burgasser2026ucdmcmc}. The input vector comprised 54 F444W-normalized spectral bins spanning 4.06-4.84~$\micron$ (effective $R\sim307$) and three flux ratios sharing F444W as the reference band: F115W/F444W, F150W/F444W, and F277W/F444W. The spectral information was compressed using principal component analysis \citep[PCA;][]{JolliffeCadima2016}, and random-forest regressors \citep{Breiman2001} were implemented using \texttt{scikit-learn} \citep{Pedregosa2011}. Spectral preprocessing, PCA, and regression were refitted within each validation fold, and each fold withheld all grid models having the target-parameter value under evaluation. 

For effective temperature, 15 interior held-out slices yielded $R^2=0.997$, an RMSE of 15.7~K, and an MAE of 12.5~K. Combining the measurement-noise Monte Carlo realizations with the empirical held-out-slice residuals gave
\begin{equation}
   T_{\rm eff}=712^{+19}_{-21}\ {\rm K}. 
\end{equation}

The corresponding interval overlaps the LOWZ MCMC estimate of $718.6^{+10.6}_{-10.2}$~K. In a separate feature-set ablation, the validation RMSE decreased from 33.9~K using photometry alone and 35.4~K using the spectrum alone to 16.8~K when both were combined, indicating that the two data components provide complementary temperature information.

For metallicity, nine interior held-out slices gave $R^2=0.988$, an RMSE of 0.097~dex, and an MAE of 0.065~dex. The resulting estimate was
\begin{equation}
    [\mathrm{M/H}]=-1.50^{+0.25}_{-0.35}.
\end{equation}
This interval does not overlap the LOWZ MCMC result of $[\mathrm{M/H}]=-1.00^{+0.07}_{-0.08}$. Nevertheless, both independently favor a subsolar atmosphere. It is not a statistically independent measurement because it uses the same observations, F444W normalization, fixed-chemistry assumptions ($\mathrm{C/O}=0.55$ and $\log K_{zz}=-1.0$), and LOWZ model grid as the MCMC analysis. The exact metallicity should therefore be regarded as method- and model-dependent.

\subsection{Molecular-opacity and empirical spectral comparisons}
\label{sec:molecular_results}

The fixed-atmosphere PICASO calculation provides a close reconstruction
of the observed F444W spectral shape. The full Sonora Elf Owl
V2/PICASO atmosphere gives
$\chi^2_{\rm spec}=1.130$ and a cosine similarity of 0.957. 
These values are close to, but not identical to, those of the
precomputed Sonora Elf Owl grid spectrum in Table~\ref{tab:results} because the
same atmospheric structure and chemistry are recomputed here
through the PICASO radiative-transfer and opacity framework for
the molecule-removal experiment.

\begin{figure*}
    \centering
    \includegraphics[width=\textwidth]{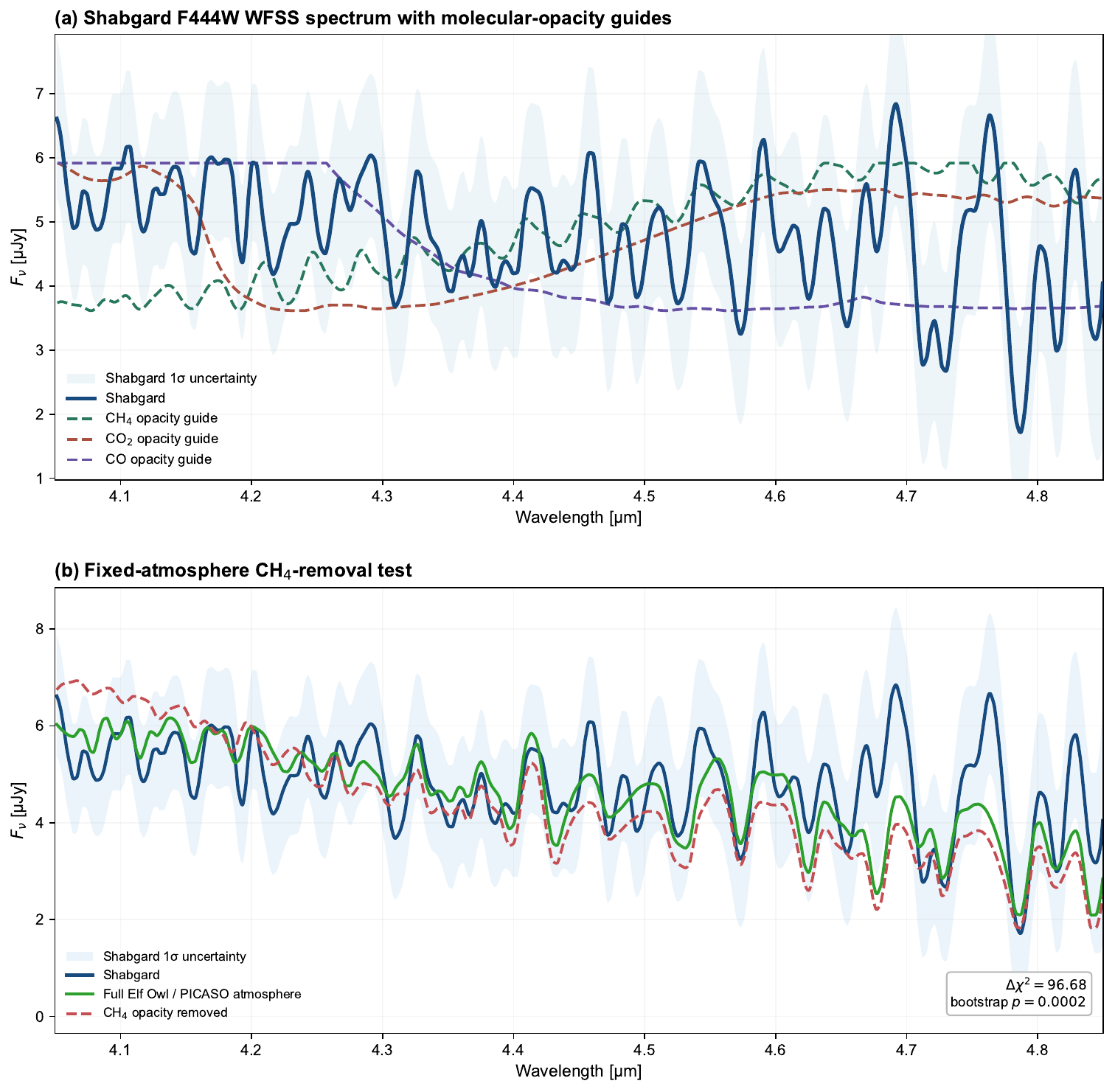}
    \caption{
    Molecular identification in the Shabgard F444W spectrum.
    (a) Shabgard is shown at its native NIRCam/WFSS resolving power
    ($R\sim1600$) together with representative ExoMolOP
    CH$_4$, CO$_2$, and CO opacity profiles displayed at
    $R\sim300$ as qualitative wavelength guides.
    The opacity curves are independently normalized, inverted, and
    rescaled and are not physical flux components.
    (b) Fixed-atmosphere CH$_4$-removal test based on the selected
    Sonora Elf Owl V2 atmosphere recomputed with PICASO.
    The green curve shows the full atmosphere and the red dashed curve
    shows the same fixed atmosphere with CH$_4$ opacity removed.
    Removing CH$_4$ worsens the fit by
    $\Delta\chi^2=96.68$, with an empirical bootstrap probability
    $p=2.0\times10^{-4}$.
    }
    \label{fig:molecular_test}
\end{figure*}
Figure~\ref{fig:molecular_test} shows the molecular-opacity comparison.
The broad ExoMolOP curves in panel (a) are shown only as qualitative
wavelength guides. In the physical leave-one-species-out calculation
shown in panel (b), removing CH$_4$ opacity substantially worsens the
fit, with
$\Delta\chi^2=96.68$ and an empirical bootstrap probability
$p=2.0\times10^{-4}$. The result is insensitive to a reasonable
change in the adopted resolving power:
$\Delta\chi^2=97.94$, 96.68, and 95.60 for
$R=1400$, 1600, and 1800, respectively.
In contrast, removing CO$_2$ produces only
$\Delta\chi^2=0.048$ ($p=0.051$), which we regard as inconclusive,
while removing CO gives $\Delta\chi^2=-5.09$ ($p=0.959$) and is not
supported by this fixed-atmosphere test.

We therefore find strong fixed-atmosphere support for CH$_4$ opacity
contributing to the observed 4-5~$\mu$m spectral shape.
We do not interpret this result as a formal CH$_4$ abundance
measurement because the atmospheric parameters are held fixed and the
available wavelength interval contains broad, overlapping molecular
bands.

\begin{figure*}
    \centering
    \includegraphics[width=\textwidth]
    {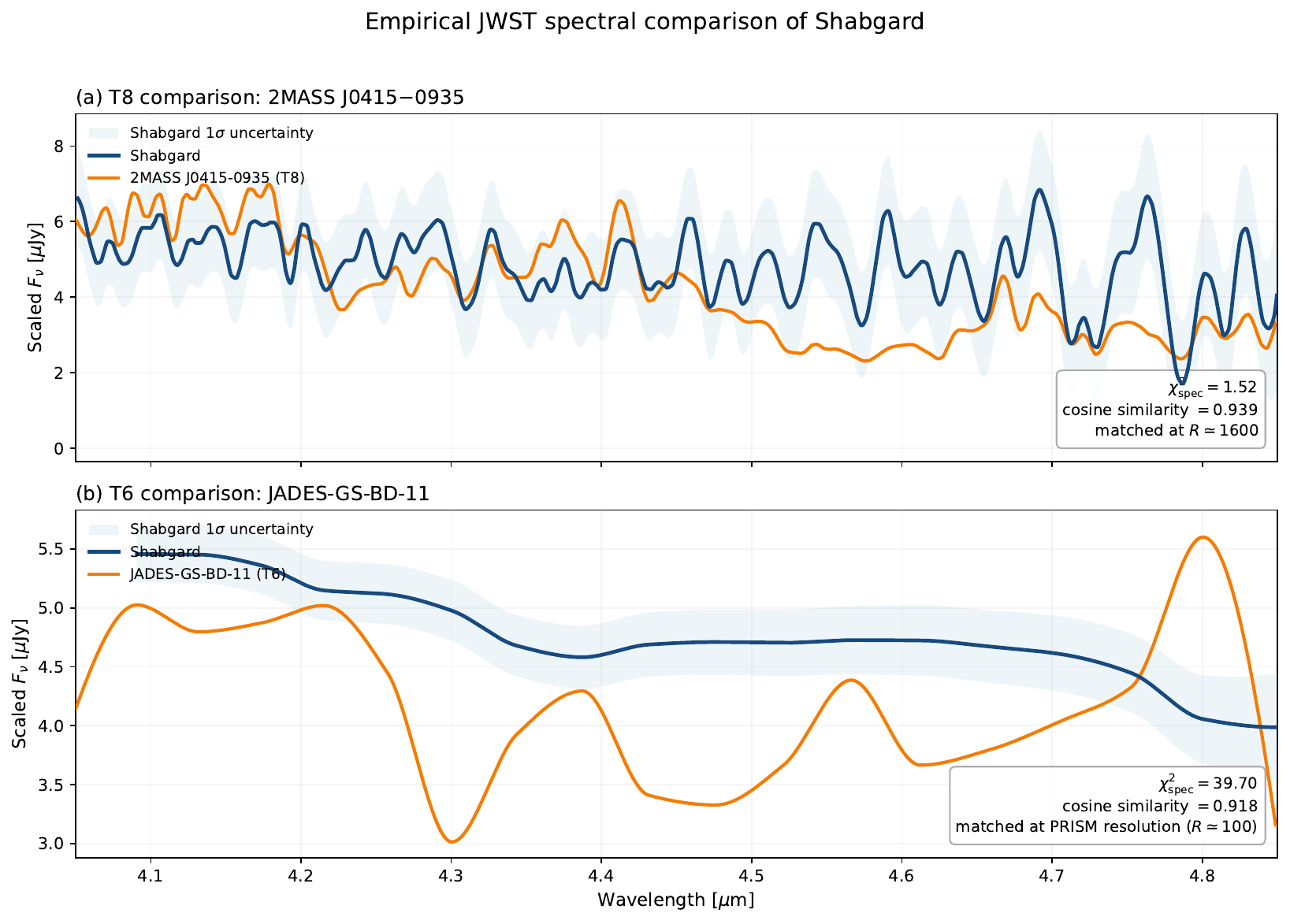}
    \caption{
    Direct empirical JWST spectral comparison of Shabgard with two
    brown dwarfs having similar effective temperatures.
    In panel (a), the higher-resolution
    2MASS~J0415$-$0935 NIRSpec/G395H spectrum is matched to the
    native Shabgard NIRCam/WFSS resolving power
    ($R\sim1600$).
    In panel (b), Shabgard is degraded to the nominal NIRSpec/PRISM
    resolving power ($R\sim100$) for comparison with
    JADES-GS-BD-11.
    A single multiplicative normalization is fitted in each comparison.
    The reported $\chi^2_{\rm spec}$ values use the same
    mean uncertainty-weighted squared-residual statistic defined in
    Section~\ref{sec:methods}, while the cosine similarity is the same
    raw, uncentered diagnostic used in the atmosphere-model comparison.
    The plotted curves are smoothed only for visualization after the
    diagnostics are calculated.
    }
    \label{fig:empirical_bd_comparison}
\end{figure*}

The empirical JWST comparison is shown in
Figure~\ref{fig:empirical_bd_comparison}.
At the native Shabgard information scale, the T8 dwarf
2MASS~J0415$-$0935 provides a close detailed match, with
$\chi^2_{\rm spec}=1.52$ and cosine similarity $=0.939$.
For JADES-GS-BD-11, the comparison is performed after degrading
Shabgard to the nominal PRISM resolution; the cosine similarity remains
high (0.918), but the substantially larger
$\chi^2_{\rm spec}=39.70$ indicates poorer wavelength-by-wavelength
agreement.

Because these two comparisons contain different numbers of independent
resolution elements, we do not use their absolute
$\chi^2_{\rm spec}$ values as a formal subtype-selection statistic.
Instead, the comparison provides an empirical spectral-shape check.
The closer agreement with 2MASS~J0415$-$0935 supports a late-T-like
interpretation of Shabgard, consistent with the approximately
700~K temperature inferred from the atmosphere models, while the
restricted F444W interval remains insufficient for a formal empirical
spectral classification.

\section{Discussion}\label{sec:discussion}

\subsection{Why the brown-dwarf interpretation is favored}

Shabgard is favored as a brown-dwarf because several independent
tests point in the same direction. The source was selected as a compact,
red, point-source-like object from the COSMOS2025/COSMOS-Web photometric
candidate catalog. It is also one of the few candidates with a centered and
relatively clean COSMOS-3D F444W WFSS trace. The local-background,
profile-weighted extraction gives a synthetic F444W flux consistent with the
COSMOS2025 catalog photometry, so the spectrum and photometry can be compared
on a common flux scale.

The atmosphere-model comparison provides the strongest physical support for the
brown-dwarf interpretation. Five independent model families converge on
$T_{\rm eff}=700$~K, while ATMO provides a cooler $450$~K comparison
solution; the LOWZ continuous posterior independently centers near $719$~K.
The best-ranked solutions favor moderate-to-high gravities. This temperature
scale is broadly associated with the late-T to Y-dwarf regime
\citep{Burgasser2006,Cushing2011,Filippazzo2015}, although the
restricted F444W wavelength coverage does not permit a formal
empirical spectral classification. These solutions are not
consistent with a hot stellar photosphere, and the compact morphology and
stellar-template preference from the photometric selection disfavor an ordinary
galaxy interpretation. A deeper JWST/NIRSpec PRISM spectrum would provide the broader wavelength coverage and higher signal-to-noise ratio required to assign a more secure empirical spectral type and to test the molecular absorption pattern expected for a late-T/Y dwarf. Complementary MIRI/LRS spectroscopy would extend this analysis into the mid-infrared, where strong molecular features, particularly from NH$_3$, could provide additional constraints on the atmospheric chemistry and effective temperature \citep{Beiler2023}. The multi-epoch astrometry provides an additional test that is independent of the atmosphere fitting. The measured displacement is reproduced in three NIRCam filters and is substantially larger than the apparent motions recovered for stationary artificial sources. This proper motion strongly disfavors a stationary high-redshift source. It does not by itself establish a brown-dwarf classification, because other Galactic point sources may also have measurable motions, but its combination with the compact morphology, very red photometry, and cool atmosphere-model solutions considerably strengthens the foreground ultracool-dwarf interpretation.

\subsection{Where the model families agree and where they differ}

The model families agree on three main points. First, Shabgard is cool: five
grids converge at $T_{\rm eff}=700$~K, ATMO remains in the cool regime at
$450$~K, and the LOWZ continuous posterior centers near $719$~K. Second, the
best-ranked solutions generally favor moderate-to-high gravity, although the
exact value remains model-dependent. Third, the inferred physical scale is
substellar, even though the exact mass differs among model assumptions.

The main differences involve metallicity, chemistry, clouds, and the mapping
from atmosphere parameters to mass and distance. The metallicity-sensitive
models favor subsolar abundance: LOWZ gives [M/H]$=-1.00$ in both the displayed
grid fit and the MCMC posterior, and SAND gives [M/H]$=-1.10$ in the
displayed comparison. Sonora Elf Owl and Sonora Bobcat also select the lowest
metallicity available in the fitted Sonora grids, [M/H]$=-0.50$. This agreement
at the low-metallicity edge is encouraging, but it should not be over-
interpreted as a precise abundance constraint. In the F444W grism range,
metallicity, C/O ratio, gravity, and vertical mixing can all affect the
continuum shape and molecular opacity \citep{Mukherjee2024}. The current data
are therefore better suited to identifying a cool, likely metal-poor atmosphere
than to measuring a unique chemical composition.

The cloud comparison is also model-dependent. The Morley grid includes sulfide
and salt condensate clouds, while Sonora Bobcat and ATMO are used here as
cloud-free or effectively cloud-free comparison models. The Morley best fit
has $f_{\rm sed}=5$, which corresponds to efficient sedimentation and a
relatively thin or settled cloud layer in the Ackerman-Marley framework
\citep{AckermanMarley2001,Morley2012}. Its best fit shows that cloud opacity
is allowed, but it does not prove that clouds are present. With only the F444W
window, the safest conclusion is that the data are consistent with either a
cloud-free atmosphere or a cloudy atmosphere with relatively settled condensate
opacity.

The physical-scale disagreement is also understandable. The ATMO and Sonora
Bobcat evolutionary checks give masses near $23\mjup$ and distances near
$145$ pc. The LOWZ atmosphere-only posterior, combined with a radius prior,
gives a higher conditional mass of $52^{+18}_{-16}\mjup$ and a distance of
$176^{+18}_{-18}$ pc. These values differ because the LOWZ calculation fits a
dilution factor for an atmosphere grid and then imposes an external radius
prior, while the ATMO/Sonora checks use evolutionary tracks that link radius,
mass, gravity, temperature, and age self-consistently. The difference should
therefore be treated as a systematic uncertainty in the physical scale rather
than as a contradiction.

\subsection{Information content of the F444W spectrum}
\label{sec:f444w_information}

The F444W grism range samples part of the 4-5~$\mu$m region in which
cool brown-dwarf spectra are shaped by molecular opacity from species
including CH$_4$, H$_2$O, CO, CO$_2$, NH$_3$, PH$_3$, H$_2$S,
and collision-induced H$_2$ absorption
\citep{Marley2021,Phillips2020,Mukherjee2024,Tennyson2012}.
This wavelength interval therefore provides useful atmospheric
information even though it is substantially narrower than a full
NIRSpec spectrum.

The qualitative ExoMolOP comparison and the fixed-atmosphere PICASO
experiment provide complementary information. The opacity guides show
where broad molecular absorption is expected, whereas the
leave-one-species-out calculation tests whether a particular opacity
source is needed to reproduce the observed spectral shape.
Within the selected Sonora Elf Owl V2 atmosphere, removing CH$_4$
produces a substantial degradation in the fit
($\Delta\chi^2=96.68$, bootstrap
$p=2.0\times10^{-4}$), and this result remains stable over
$R=1400$-1800. We therefore find strong fixed-atmosphere support for
CH$_4$ opacity in the observed F444W spectrum.
The corresponding tests do not provide comparable evidence for
CO$_2$ or CO.

This result should nevertheless not be interpreted as a molecule-by-molecule
abundance retrieval or as the identification of isolated CH$_4$
transitions. The molecular bands in this wavelength interval are broad
and overlapping, and the PICASO experiment holds the pressure-temperature
profile and remaining chemistry fixed. Broader wavelength coverage is
required to marginalize simultaneously over temperature, gravity,
metallicity, C/O ratio, vertical mixing, clouds, and molecular
abundances.

The empirical comparison provides an independent check on the overall
spectral shape. Shabgard shows substantially closer detailed agreement
with the T8 dwarf 2MASS~J0415$-$0935 than with the T6
JADES-GS-BD-11 spectrum after the appropriate pairwise resolution
matching. This supports a late-T-like interpretation, but the limited
F444W wavelength range remains insufficient for a formal empirical
spectral subtype.

A deeper JWST/NIRSpec spectrum would extend the comparison to multiple
H$_2$O and CH$_4$ bands and provide the wavelength leverage needed for
a more complete atmospheric retrieval. Complementary MIRI spectroscopy
would further constrain the cool-atmosphere chemistry, particularly
through NH$_3$ and other mid-infrared molecular features.

\subsection{Distance, kinematics, and follow-up}

The present distance estimates are $\sim145$-$176$ pc, depending on whether
the physical scale is inferred from ATMO/Sonora evolutionary checks or from the
LOWZ dilution factor with a radius prior. At these distances, Shabgard would
be a foreground Galactic ultracool dwarf in the COSMOS field. The inferred mass
is more model-dependent: ATMO/Sonora evolutionary checks favor
$\sim23\mjup$, while LOWZ plus a $1\rjup$ radius prior gives
$\sim52\mjup$. Both values are below the hydrogen-burning limit and remain
within the brown-dwarf regime.

The most useful follow-up would be broader JWST/NIRSpec or combined
NIRCam+MIRI spectroscopy. A wider wavelength baseline would test the H$_2$O and
CH$_4$ bands outside the F444W window, improve the measurement of the
$4$-$5\mu\mathrm{m}$ molecular structure, and help separate metallicity, C/O ratio,
surface gravity, vertical mixing, and cloud opacity. Such data would also allow
a more secure empirical late-T/Y spectral classification.

\section{Conclusions}\label{sec:conclusions}

We have presented a spectrophotometric analysis of Shabgard
(COSMOS2025 ID~639188), a cool brown-dwarf identified in JWST/COSMOS-3D F444W WFSS data.
The main conclusions are:

\begin{enumerate}

\item The atmosphere-model comparison strongly favors a cool brown-dwarf
atmosphere. Sonora Elf Owl, LOWZ, SAND, Morley, and Sonora Bobcat all
select $T_{\rm eff}=700$~K, while ATMO provides a cooler
$T_{\rm eff}=450$~K comparison solution. The convergence of five independent
grids at $700$~K, together with the LOWZ continuous posterior at
$718.6^{+10.6}_{-10.2}$~K, makes approximately $700$~K the principal
atmospheric temperature scale of the analysis.

\item The displayed grid solutions span $\log g=4.0$-$5.5$, with most of the
highest-ranked fits favoring moderate-to-high gravity. None requires an
extremely low-gravity, inflated atmosphere. This is consistent with
Shabgard being a field brown dwarf rather than a very young object.

\item The metallicity-sensitive grids favor subsolar metallicity. LOWZ gives
[M/H]$=-1.00$ in the displayed grid fit and
[M/H]$=-1.00^{+0.07}_{-0.08}$ in the PCA-interpolated MCMC posterior, while
SAND gives [M/H]$=-1.10$ in the displayed comparison. We treat this as
evidence for a likely metal-poor atmosphere, but not as a final abundance
measurement because the F444W-only spectrum leaves metallicity, C/O ratio,
gravity, and vertical mixing partially degenerate.

\item The Morley cloudy model gives a competitive fit with $f_{\rm sed}=5$.
This corresponds to a cloudy but relatively settled/thinner-cloud solution
within the Ackerman-Marley/Morley framework. Cloud opacity is therefore not
ruled out, but the present data do not provide a direct cloud detection.

\item The physical scale remains model-dependent but substellar. ATMO/Sonora
evolutionary checks give distances near $143$-$145$ pc and masses near
$23\mjup$, while the LOWZ dilution factor combined with a
$1.00\pm0.10\rjup$ radius prior gives
$d=176^{+18}_{-18}$ pc and $M=52^{+18}_{-16}\mjup$.

\item Multi-epoch NIRCam astrometry gives consistent proper-motion
amplitudes of $88.6$-$90.9\ {\rm mas\,yr^{-1}}$ in F115W, F200W,
and F356W, with a combined value of
$|\mu|=89.17\pm0.40\ {\rm mas\,yr^{-1}}$. Artificial-source tests
show that the displacement is substantially larger than the apparent
motions recovered for stationary sources. Shabgard is therefore
securely identified as a Galactic source rather than a
stationary extragalactic object.

\item Combining the proper motion with the model-dependent distances
gives a heliocentric tangential velocity of approximately
$60$-$74\ {\rm km\,s^{-1}}$. This is elevated relative to the mean
local ultracool-dwarf population and is consistent with an older
Galactic-disk object, although a radial-velocity measurement is needed
to calculate the full $(U,V,W)$ motion and test thin-disk, thick-disk,
or halo membership quantitatively.

\item A PCA-random-forest regression on the fixed-chemistry LOWZ grid provides an independent-method cross-check of the atmospheric inference. It yields $T_{\rm eff}=712^{+19}_{-21}$~K, with an interval overlapping the LOWZ MCMC result, and independently supports a subsolar atmosphere with $[\mathrm{M/H}]=-1.50^{+0.25}_{-0.35}$. Because the regression and MCMC metallicity intervals do not overlap, the precise metallicity remains method- and model-dependent. 

\item

A fixed-atmosphere Sonora Elf Owl V2/PICASO test provides strong
support for CH$_4$ opacity in the F444W spectrum:
removing CH$_4$ worsens the fit by
$\Delta\chi^2=96.68$, with an empirical bootstrap probability
$p=2.0\times10^{-4}$. Comparable support is not obtained for
CO$_2$ or CO. An independent empirical comparison shows that
Shabgard has closer detailed spectral-shape agreement with the T8
dwarf 2MASS~J0415$-$0935 than with the T6 JADES-GS-BD-11 spectrum,
supporting a late-T-like interpretation without assigning a formal
spectral subtype.

\end{enumerate}

Taken together, the compact morphology, red photometry, cool
atmosphere-model solutions, and significant proper motion identify
Shabgard as a strong foreground cool brown-dwarf, likely
with a late-T/Y-like atmosphere and subsolar metallicity.
Higher-signal-to-noise and broader-wavelength spectroscopy,
particularly with JWST/NIRSpec and potentially MIRI, is required to
assign an empirical spectral type, identify individual molecular
features, and better constrain its metallicity, chemistry, and cloud
properties.

\begin{acknowledgments}
We thank the referee for the constructive feedback that improved the paper.
This research is based on observations made with the NASA/ESA/CSA James Webb
Space Telescope. The data were obtained from the Mikulski Archive for Space
Telescopes (MAST) at the Space Telescope Science Institute, which is operated
by the Association of Universities for Research in Astronomy, Inc. (AURA),
under NASA contract NAS 5-03127. This work uses JWST/NIRCam imaging from the
COSMOS-Web program and JWST/NIRCam wide-field slitless spectroscopy from the
COSMOS-3D program (GO-5893; PI: K. Kakiichi). We thank the COSMOS-Web and
COSMOS-3D teams for producing and making available the data products that made
this analysis possible.

We thank the COSMOS-3D team for their work on the
NIRCam/WFSS reduction products and for helpful clarification on the extraction
and flux-calibration treatment. We also thank the brown-dwarf model and
software communities whose public grids and tools made the atmosphere-model
comparison possible, including SAND, Sonora, Morley, ATMO, LOWZ, and
\texttt{ucdmcmc}.
This work also uses JWST/NIRCam imaging from the Public Release
IMaging for Extragalactic Research program
(PRIMER; GO-1837; PI: J. Dunlop).

T.G. acknowledges the support of the National Science and Technology Council
of Taiwan through grants 114-2112-M-007-026, 113-2112-M-007-006, and
113-2927-I-007-501. T.H. acknowledges the support of the National Science and
Technology Council of Taiwan through grants 110-2112-M-005-013-MY3 and
110-2112-M-007-034. RAM acknowledges support from the Swiss National Science Foundation (SNSF) through project grant 200020\_207349.

We also thank our collaborators and colleagues for valuable discussions and
feedback that improved this work.

\end{acknowledgments}

\software{
Astropy (v7.2.0; \citealt{Astropy2022}),
Astroquery (v0.4.11),
NumPy (v2.4.6; \citealt{Harris2020}),
SciPy (v1.17.1; \citealt{Virtanen2020}),
pandas (v3.0.3),
Matplotlib (v3.10.9; \citealt{Hunter2007}),
Photutils (v3.0.0),
reproject (v0.19.0),
h5py (v3.15.1),
xarray (v2026.4.0),
netCDF4 (v1.7.4),
scikit-learn (v1.9.0; \citealt{Pedregosa2011}),
PICASO (v4.0.1; \citealt{Batalha2019}),
emcee (v3.1.6; \citealt{ForemanMackey2013}),
corner (v2.2.3; \citealt{ForemanMackey2016}),
\texttt{ucdmcmc} (v1.4.4),
the \texttt{unfold\_jwst} reduction framework,
and the JWST Calibration Pipeline (v1.20.2).
}

\section*{Data Availability}

The JWST observations used in this work are publicly available through
the Mikulski Archive for Space Telescopes (MAST). The analysis uses
COSMOS-Web imaging (GO-1727), COSMOS-3D spectroscopy (GO-5893), and
PRIMER imaging (GO-1837). The COSMOS2025 catalog is publicly available
through the COSMOS2025 data release. Derived products supporting this
study are available from the corresponding author upon reasonable
request.

\clearpage
\appendix
\setcounter{figure}{0}
\renewcommand{\thefigure}{A\arabic{figure}}

\section{Additional diagnostic and physical-scale checks}
\label{app:additional_checks}

\subsection{Extraction diagnostics for the other centered-trace candidates}
\label{app:other_extractions}

As described in Section~\ref{sec:extraction}, the one-dimensional spectrum of
Shabgard was obtained through a custom local-background,
profile-weighted extraction from the COSMOS-3D
\texttt{unfold\_jwst} two-dimensional product. Before selecting
Shabgard for the detailed atmosphere-model analysis, we inspected
the available two-dimensional and one-dimensional products for all
photometrically selected candidates with COSMOS-3D coverage. Six
objects had traces that were sufficiently centered and isolated to
permit an attempted local-background extraction.

\begin{figure*}[ht!]
\centering
\includegraphics[width=0.95\textwidth]
{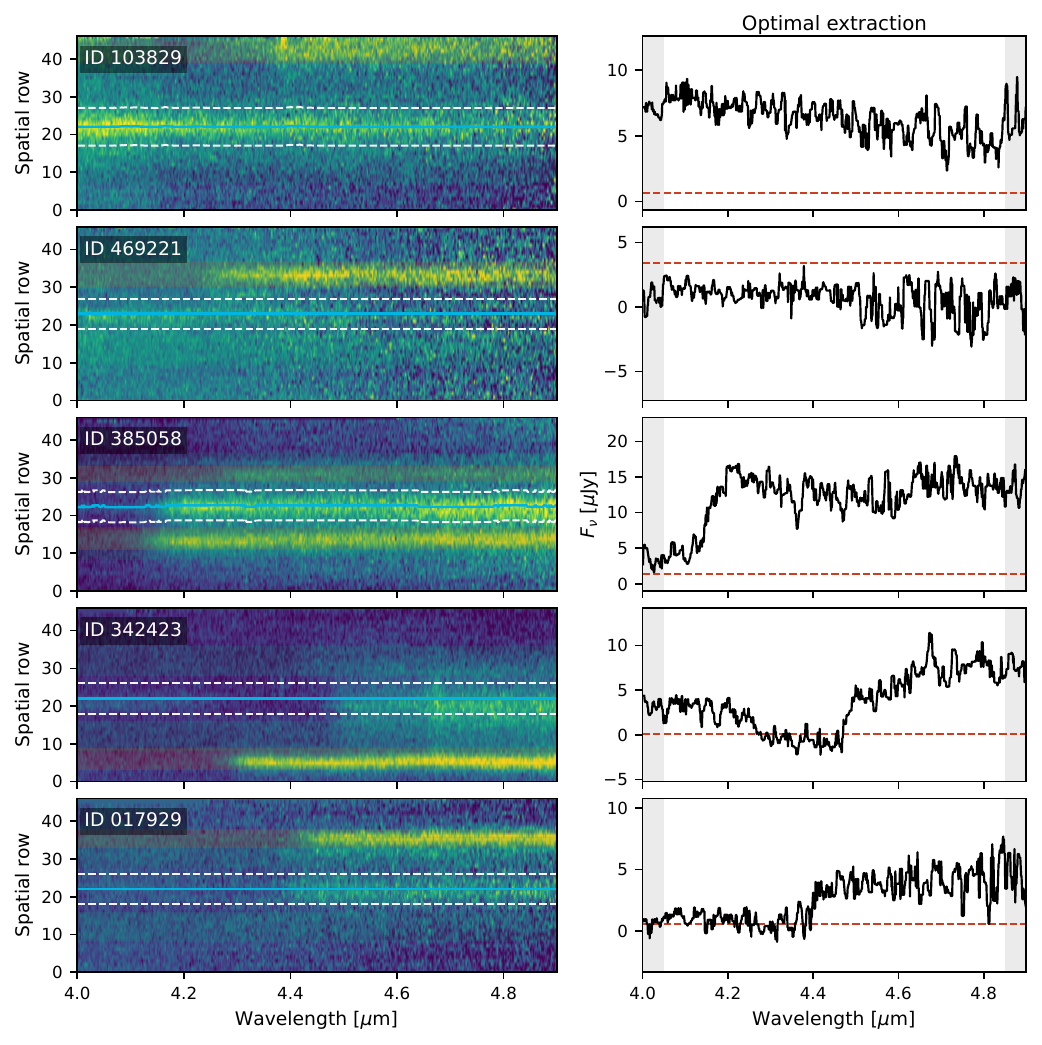}
\caption{
Local-background extraction diagnostics for the other five
centered-trace candidates. For each object, the left panel shows the
two-dimensional COSMOS-3D F444W WFSS spectrum, with the traced source
position and adopted extraction regions marked, while the right panel
shows the resulting local-background-subtracted, profile-weighted
one-dimensional spectrum. These objects were not included in the
detailed atmosphere-model comparison because their extractions were
less stable than that of Shabgard owing to lower continuum signal,
residual spectral overlap, background sensitivity, or weaker
agreement with the independent F444W photometry. The main physical
analysis in this paper therefore focuses on Shabgard.
}
\label{fig:other_extractions}
\end{figure*}

Figure~\ref{fig:other_extractions} shows the extraction diagnostics
for the remaining five centered-trace candidates. The same general
procedure used for Shabgard was applied to these objects: the source
trace was identified in the two-dimensional coadd, local background
regions were defined on both sides of the trace, and a
profile-weighted one-dimensional spectrum was extracted after
background subtraction.

These five candidates were not carried forward to the detailed
atmosphere-model comparison because their extracted spectra were less
suitable for quantitative fitting. Depending on the object, the main
limitations included a weaker continuum signal, residual structure
from neighbouring dispersed sources, sensitivity to the adopted
background regions, or poorer agreement between the extracted F444W
flux scale and the independent COSMOS2025 photometry. The selection of
Shabgard was therefore based on the stability and quality of the
spectral extraction before consideration of the atmosphere-model fit
results. The spectra shown here are retained as extraction and
candidate-selection diagnostics and should not be interpreted as
ruling out a brown-dwarf nature for the other candidates.

\clearpage
\subsection{Evolutionary physical-scale checks}
\label{app:evolution_checks}

The conversion of an atmosphere-model normalization into physical
properties depends on the adopted relation among effective
temperature, surface gravity, radius, mass, and age. We therefore
performed supplementary physical-scale checks using the ATMO 2020
and Sonora Bobcat evolutionary calculations
\citep{Phillips2020,Marley2021}. These checks provide an independent,
model-dependent comparison with the LOWZ dilution-factor analysis
presented in Section~\ref{lowz_mcmc}.

\begin{figure*}[t]
\centering
\includegraphics[width=0.98\textwidth]
{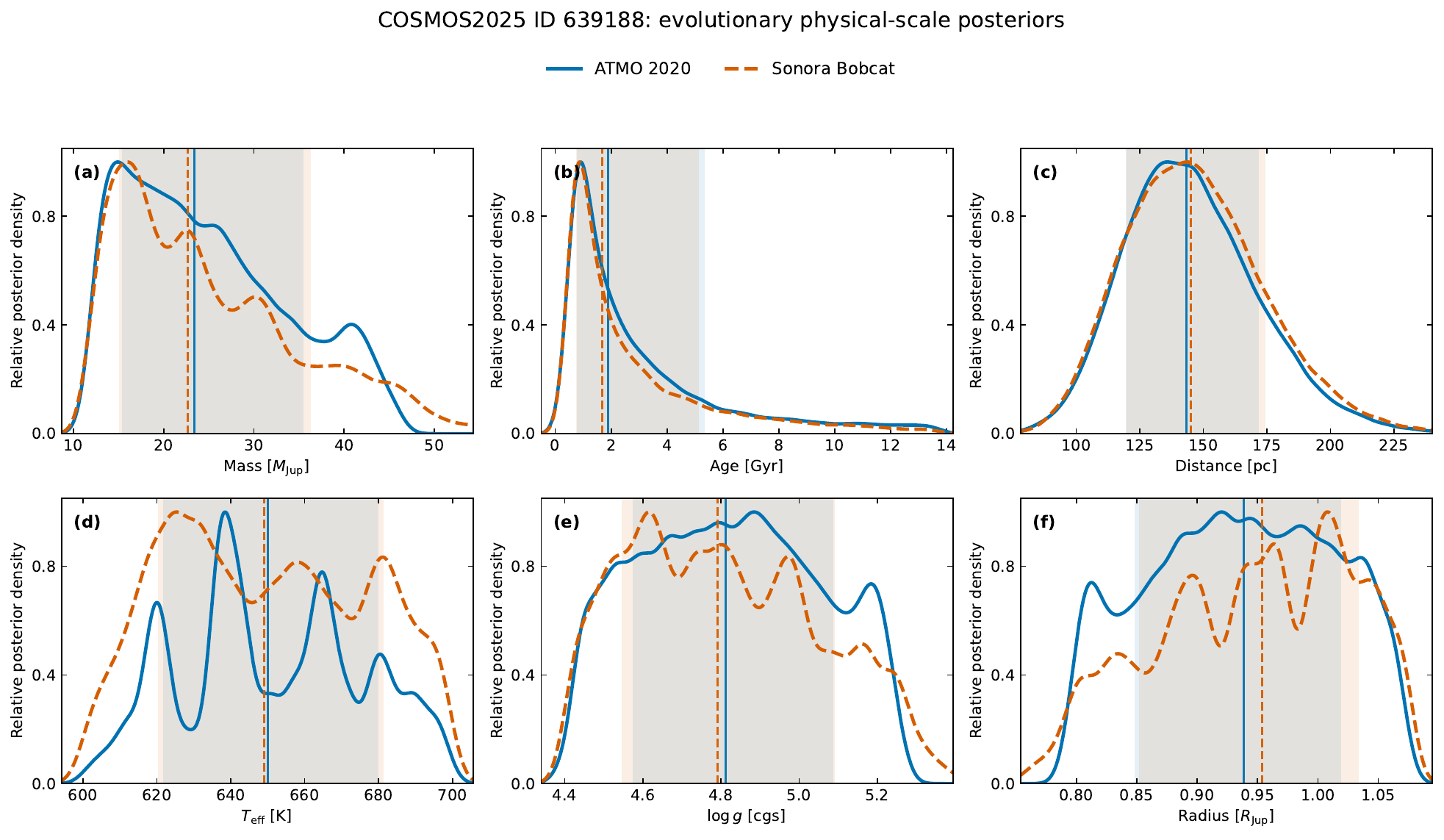}
\caption{
Marginal posterior distributions of the evolutionary physical
properties of Shabgard obtained using the ATMO 2020 and Sonora
Bobcat model families. The panels show (a) mass, (b) age,
(c) distance, (d) effective temperature, (e) surface gravity, and
(f) radius. Solid blue curves represent ATMO 2020, while dashed
orange curves represent Sonora Bobcat. The vertical lines mark the
posterior medians, and the shaded regions indicate the corresponding
16th-84th percentile intervals. Each distribution is independently
normalized to its maximum for visual comparison. Both evolutionary
frameworks favor a mass near $23\,M_{\mathrm{Jup}}$, a distance near
$145$~pc, an effective temperature near $650$~K, a surface gravity
near $\log g=4.8$, and a radius close to one Jupiter radius. The
multimodal structure visible in some derived parameters reflects the
discrete and nonlinear mapping through the evolutionary-model grids.
}
\label{fig:evol}
\end{figure*}

For each evolutionary framework, posterior samples were constructed
over an age range of $0.5$-$13.8$~Gyr. The atmosphere solution was
linked to the corresponding evolutionary-model mass, radius,
effective temperature, and surface gravity. The model normalization
was then converted into a distance using the flux-dilution relation
\begin{equation}
    \alpha = \left(\frac{R}{d}\right)^2,
\end{equation}
where $R$ is the evolutionary-model radius and $d$ is the distance.
Figure~\ref{fig:evol} shows the resulting marginal posterior
distributions.

The ATMO 2020 calculation gives a mass of
$23.4^{+12.0}_{-8.1}\,M_{\mathrm{Jup}}$, a distance of
$143^{+28}_{-24}$~pc, and a radius of
$0.94^{+0.08}_{-0.09}\,R_{\mathrm{Jup}}$. The corresponding Sonora
Bobcat values are
$22.7^{+13.6}_{-7.6}\,M_{\mathrm{Jup}}$,
$145^{+29}_{-25}$~pc, and
$0.95^{+0.08}_{-0.10}\,R_{\mathrm{Jup}}$. The two frameworks also
give consistent effective temperatures and surface gravities, with
median values near $650$~K and $\log g\simeq4.8$, respectively.

The close agreement between the two evolutionary calculations
indicates that the inferred physical scale is not strongly dependent
on the choice between these model families. Nevertheless, the
posteriors remain conditional on the atmosphere fits, evolutionary
tracks, adopted age range, and flux normalization. They should
therefore be interpreted as model-dependent consistency checks rather
than direct measurements of the mass, radius, age, or distance.
Together with the LOWZ dilution-factor estimate, these results support
the interpretation of Shabgard as a foreground substellar object at
a distance of order $150$~pc.

\bibliographystyle{aasjournalv7}
\bibliography{references}

\end{document}